\documentclass[prl,amsmath,amssymb,twocolumn,superscriptaddress]{revtex4-1}
\usepackage{graphicx,bm,color,appendix}
\usepackage[colorlinks,bookmarks=true,citecolor=blue,linkcolor=blue,urlcolor=blue]{hyperref}
\usepackage{times}

\newcommand{\rr}{{\bf r}}
\newcommand{\mr}[1]{ \mathrm{#1} }

\usepackage{appendix}

\usepackage{epstopdf}
\usepackage{xcolor}
\definecolor{orange}{rgb}{1,0.5,0}

\begin{document}
	
	\title{
		Modular transformation and anyonic statistics of multi-component fractional quantum Hall states}

	\author{Liangdong Hu}
	\affiliation{	School of Physics, Zhejiang University, Hangzhou 310058, China}
	
	\affiliation{Westlake Institute of Advanced Study,	 Westlake University, Hangzhou 310024, China }
	
	\author{Zhao Liu}
	\email{zhaol@zju.edu.cn}
	\affiliation{	School of Physics, Zhejiang University, Hangzhou 310058, China}
	\affiliation{Zhejiang Institute of Modern Physics, Zhejiang University, Hangzhou 310027, China}

	\author{W. Zhu}
	\email{zhuwei@westlake.edu.cn}
	\affiliation{Westlake Institute of Advanced Study, Westlake University, Hangzhou 310024, China }

	\begin{abstract}
		We investigate the response to modular transformations and the fractional statistics of Abelian multi-component fractional quantum Hall (FQH) states. In particular, we analytically derive the modular matrices encoding the statistics of anyonic excitations for general Halperin states using the conformal field theories (CFTs). 
		We validate our theory by several microscopic examples, including the spin-singlet state using anyon condensation picture and the Halperin (221) state in a topological flat-band lattice model using numerical calculations. 
		Our results, uncovering that
		the modular matrices and associated fractional statistics are solely determined by the $K$-matrix, further strengthens the correspondence between the 2D CFTs and (2+1)D topological orders for multi-component FQH states.  
	\end{abstract}
	
	\date{\today}
	
	\maketitle

\begin{figure}[t]
		\centering
		\includegraphics[width=\linewidth]{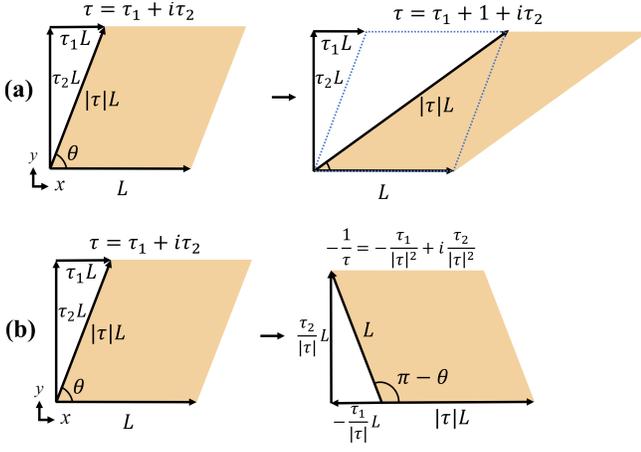}
		\caption{\label{fig:modular_ST}
			The torus geometry is defined by two fundamental vectors $\vec L_2=L\vec \tau$ and $\vec L_1=L \vec e_x$, 
			and the twist angle is $\theta$. 
			(a) The $\mathcal{T}$ transformation sends $\vec \tau=\tau_1 \vec e_x+\tau_2 \vec e_y$ to its equivalent geometry $\vec \tau+\vec e_x$,
			thus leaving the torus geometry unchanged. 
			(b) The modular $\mathcal S$ transformation generates a counterclockwise rotation and transforms the torus spanned by $L \vec{e}_x$ and $\vec{\tau} L$ to a torus spanned by $|\tau| L \vec{e}_x$ and $-\frac{|\tau|L}{\vec \tau}$. 
		}
	\end{figure}

	\clearpage
	\textit{Introduction.---}
	Fractionalized quasiparticles~\cite{Laughlin83,MOORE1991362,RevModPhys.80.1083,Feldman_2021} is a defining feature of the topological orders in fractional quantum Hall (FQH) states. These quasiparticles, dubbed anyons, obey exotic fractional statistics~\cite{Leinaas1977,PhysRevLett.48.1144} which emerges only in two spatial dimensions. How to effectively describe and systematically classify the fractional statistics and the associated topological orders is a key problem in the study of the FQH effect.
	On the one hand, Abelian-type anyonic statistics has been well captured by the Chern-Simons effective field theories describing the low-energy properties in the bulk of the FQH states~\cite{Wen1995,Wen1992a,Wen_book}. 
	On the other hand, the wave functions of many FQH states, including those with quasiparticle excitations, are shown to be conformal blocks of suitable rational conformal field theories (CFTs) ~\cite{yellowbook,MOORE1991362,Read1999,Hansson2017,Crepel2018}. As a result, the fractional statistics of FQH quasiparticles is formally encoded in the modular matrices~\cite{VERLINDE1988360,Wen1990,Rowell2009} which contain the information of mutual and self statistics, quantum dimensions, and the fusion rule of quasiparticles. For a specific FQH state, the modular matrices can be defined 
	by the degenerate ground states in response to modular transformations (see Fig. \ref{fig:modular_ST}). 
	Indeed, a large class of Abelian~\cite{YiZhang2012,WZhu2013,Liu2019} and non-Abelian~\cite{WZhu2014} FQH states has been successfully classified by their modular matrices.
	
	So far the identification of FQH states via the modular matrices is mostly limited to single-component systems~\cite{YiZhang2012,WZhu2013,WZhu2014,Liu2019}. When particles possess more internal degrees of freedom (e.g. spins, valleys, layers), the scope of FQH physics further expands to the multi-component case~\cite{Girvin,Halperin1983}. Due to the great variety and tunability of effective interactions, multi-component FQH states provide a playground for realizing emergent topological orders that have no analogue in single-component systems. In this context, how to generalize the modular-matrix approach to multi-component FQH states is largely unexplored before. Furthermore, the $K$-matrix formalism plays a crucial role in the bulk description of Abelian multi-component FQH states~\cite{Wen1995,Wen1992a,Wen_book}. It is thus natural to ask whether the CFT-FQH correspondence~\cite{Witten1989,Hansson2017} can be supported by relating modular matrices extracted from the CFT to the $K$ matrix for multi-component states. 
	
	In this paper, we address these questions critically. First, we analytically derive the modular matrices of general Abelian multi-component Halperin states~\cite{Halperin1983} directly from the underlying CFT. Remarkably, the result establishes an explicit relation between the modular matrices and the $K$ matrix in the corresponding Chern-Simons theory. Moreover, we reach a consistent conclusion by considering the response of the Halperin bulk wave functions to modular transformations. Our result thus provides compelling evidence for the correspondence between the 2D CFTs and (2+1)D bulk topological orders for multi-component FQH states. 
	We support our theory by independently deriving modular matrices of the spin-singlet Halperin states using the anyon condensation picture and using extensive exact diagonalization in a microscopic lattice model.
	
	\textit{Modular matrices from CFT.---}
	Modular matrices capture the quasiparticle statistics of topologically ordered states~\cite{Wen1990,Wen1993,Wen1995,KITAEV20062}. To be specific, 
	the modular $\mathcal S$-matrix contains braiding statistics and the fusion rules of anyons, and 
	the modular $\mathcal T$-matrix contains the self statistics of quasiparticles. These two matrices are respectively related to the modular $\mathcal S$ and $\mathcal T$ transformation on a torus (Fig.~\ref{fig:modular_ST}) \cite{YiZhang2012}. 
	
	Our goal in this section is to calculate the modular $\mathcal S$ and $\mathcal T$ matrices for Abelian multi-component FQH states using CFT, which was rarely studied before to our best knowledge. For simplicity  we only consider the bosonic states throughout this paper. Recall that the single-component bosonic Laughlin state at filling $\nu=1/(2m)$ is described by the compactified boson which manifests the $\widehat{u(1)}_{2m}$ CFT (see Appendix~I.1)~\cite{MOORE1991362,Dong2008,Wen_book,Wen1990,Wen1991,Wen1992}.  
	To incorporate Abelian multi-component FQH states, one can generalize the $\widehat{u(1)}$ theory to $\widehat{u(1)}_{\kappa,K}$ \cite{U1_level_k,Wen1992}, where the level is no longer a single number but a $\kappa\times \kappa$ positive-definite integer matrix $K$. It has been proposed that 
	many physical quantities (e.g. fractional charge, Hall conductance) of Abelian multi-component FQH states can be well captured by this $K$ matrix~\cite{Wen1992,Wen_book}.

\begin{figure}[t]
			\centering
			\includegraphics[width=\linewidth]{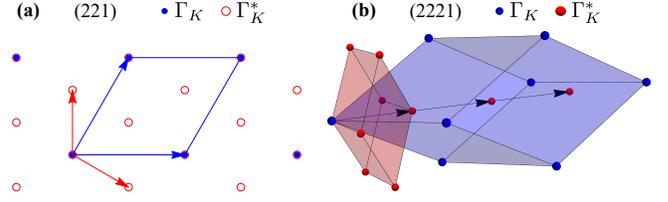}
			\caption{\label{fig:Klattice}
				Schematic plot of the $K$-lattice and its dual lattice. The blue dots and red dots respectively denote the $\Gamma_K$ lattice spanned by $\{\bm e_J\}$ and the $\Gamma_K^*$ lattice spanned by $\{\bm e_J^*\}$. The coset $\Gamma^*_K/\Gamma_K$ is the parallelogram
				spanned by $\{\bm e_J\}$ (shaded by light blue). Here we draw two examples of $\Gamma_K$ and $\Gamma_K^*$ for (a)  the Halperin $(221)$ state with      $K=\begin{pmatrix}        2 & 1\\ 1 & 2     \end{pmatrix}$ and (b) the Halperin $(2221)$ state. 
			}
		\end{figure}

	In this context, the partition function of the multi-component state can be written as summation of the character of various topological sectors $\bm a$:
	\begin{equation}\begin{split}
			Z(K) &= \sum_{\bm a \in \Gamma^*_K/\Gamma_K} |\chi_{\bm a}(\tau)|^2, 
	\end{split}\end{equation}
	where the character $\chi$ is expressed by $q$-expansion as
	\begin{equation}\begin{split}
			\chi_{\bm a}(\tau) &= \frac{1}{\eta(\tau)^\kappa}\sum_{\bm n\in\Gamma_K}q^{\frac12 (\bm n+\bm a)\cdot (\bm n+\bm a)}
	\end{split}\end{equation}
	with $q=e^{2\pi\tau i}$. Here the vector $\vec \tau$  parameterizes  the torus (Fig.~\ref{fig:modular_ST}) and  $\eta$ is Dedekind's function.
	We formulate different topological sectors by the so-called $K$-lattice $\Gamma_K$ and its dual lattice $\Gamma_K^*$, as shown in Fig.~\ref{fig:Klattice}.  $\Gamma_K$ denotes a set of vectors $\{\bm n=\sum_{I=1}^\kappa n_I \bm e_I|n_I\in\mathbb{Z} \}$, with the basis satisfying $\bm e_I\cdot\bm e_J=K_{IJ}$ ($K_{IJ}$ the element of K matrix). The corresponding dual lattice $\Gamma_K^*$ is spanned by the dual basis $\bm e^*_I$ satisfying 
	the relation $\bm e_I\cdot \bm e^*_J=\delta_{IJ}$.  The linear combination of dual basis $\sum_J K_{IJ}\bm e_J^*$ is exactly the basis of $\Gamma_K$: 
	$ \bm e_I \cdot \bm e_J=\sum_N K_{IN}\bm e_N^*\cdot\bm e_J=K_{IJ} $. With the help of this lattice representation, it is straightforward to express the inequivalent topological sectors as the coset of $\bm{a}\in \Gamma^*_K/\Gamma_K$~\cite{yellowbook}.

	
	The modular $\mathcal{S}$ and $\mathcal{T}$ matrices are determined by the changes of the character under respective modular transformation. For the modular $\mathcal{S}$ transformation $\tau \rightarrow -1/\tau$ as shown in Fig.~\ref{fig:modular_ST}(b), the character becomes (see Appendix~I.2 for details)
	\begin{align}\label{chi88}
			&	\chi_{\bm a}(-\frac{1}{\tau})= \frac{1}{(-i\tau)^{\kappa/2}\eta(\tau)^\kappa}\sum_{\bm n\in\Gamma_K}e^{-\frac{\pi i}{\tau}\left(\bm n\cdot \bm n+2\bm n\cdot \bm a+\bm a\cdot\bm a \right) }  \nonumber \\
			&=\frac{1}{\sqrt{|\det K|}}\frac{1}{\eta(\tau)^\kappa}\sum_{\bm m\in\Gamma_K^*}e^{\pi i \tau \bm m\cdot \bm m-2\pi i \bm m\cdot \bm a}, 
		\end{align}
		where we have used the generalized Poisson's resummation formula 
		\begin{equation}\nonumber
			\sum_{\bm q\in\Gamma_K}e^{-\pi a \bm q^2 + \bm q \cdot \bm b} = \frac{1}{\sqrt{|\det K|}a^{\frac{\kappa}{2}}}\sum_{\bm p\in \Gamma^*_K}e^{-\frac{\pi}{a}(\bm p+\frac{\bm b}{2\pi i})^2}.
		\end{equation}
		As any vector in $\Gamma_K^*$ can be expressed as 
		$\bm m=\sum_I m_I \bm e_I^*=\sum_I\left( \sum_J n_J K_{JI} + b_I \right){\bm e}_I^* =\bm n+\bm b $, 
		with $\bm n=\sum_I n_I \bm e_I\in \Gamma_K$ and $\bm b=\sum_I b_I \bm e^*_I\in \Gamma_K^*/\Gamma_K$, we can rewrite the summation over $\Gamma_K^*$ in Eq.~(\ref{chi88}) and obtain
		\begin{align}
			\chi_{\bm a}(-\frac{1}{\tau})=\sum_{\bm b\in \Gamma_K^*/\Gamma_K}\frac{e^{-2\pi i \bm a\cdot\bm b}}{\sqrt{|\det K|}} \chi_{\bm b}(\tau) \equiv 
			\sum_{\bm b\in \Gamma_K^*/\Gamma_K} \mathcal{S}_{ab} \chi_{\bm b}(\tau) , \nonumber 
		\end{align}
		which gives the modular $\mathcal{S}$ matrix as
		\begin{equation}\label{eq:modularS1}
			\mathcal{S} = \frac{e^{-2\pi i \bm a\cdot\bm b}}{\sqrt{|\det K|}}.
		\end{equation}
		Similarly, the character changes as 
		\begin{equation}
			\chi_{\bm a}(\tau+1) = e^{2\pi i(\frac12 \bm a \cdot \bm a-\frac{\kappa}{24})}\chi_{\bm a}(\tau)
		\end{equation}
		under the modular $\mathcal T$ transformation $\tau\rightarrow \tau+1$ [Fig.~\ref{fig:modular_ST}(a)], leading to modular $\mathcal{T}$ matrix
		\begin{align}\label{eq:modularT1}
			\mathcal T =e^{2\pi i(\frac12 \bm a \cdot \bm a-\frac{\kappa}{24})}.
		\end{align}

		Eqs.~(\ref{eq:modularS1}) and (\ref{eq:modularT1}) are the main results of this work. They give the exact forms of modular matrices of general multi-component bosonic Halperin states, which are
		universal and capture the global statistical features of anyons in the (2+1)D topological orders. In particular, the form of the $\mathcal S$ matrix demonstrates a clear relation with the $K$ matrix. First, the prefactor is proportional to $\sqrt{|\det K|}$, reflecting the Abelian nature of the state. Second, the braiding phases of anyons are uniquely determined by $e^{-2\pi i \bm a\cdot\bm b}$, where vectors $\bm a, \bm b$ belong to coset lattice defined by the $K$ matrix. Since the $K$ matrix plays a crucial role in the Chern-Simons theory of Halperin states, our result strengthens the correspondence between CFT and Chern-Simons descriptions of multi-component FQH states.   
		
		\textit{Concrete examples.---} Based on the results in Eqs.~(\ref{eq:modularS1}) and (\ref{eq:modularT1}), we now present the modular matrices for some typical Halperin states. First, let us consider the two-component spin-singlet Halperin $(m,m,m-1)$ state at filling $\nu=\frac{2}{2m-1}$. In this case, the $K$ matrix is 
		$\begin{pmatrix}
			m & m-1\\ m-1 & m
		\end{pmatrix}$
		and the coset $\Gamma^*_K/\Gamma_K$ contains $|\det K|=2m-1$ independent vectors $\bm a=\{ \frac{1}{2m-1}(a\bm e_1+a\bm e_2)|a=0,1,\cdots,2m-2 \}$. These $\bm a$ vectors form a one-dimensional lattice [see Fig.~\ref{fig:Klattice}(a) for $m=2$]. Accordingly, we have
		\begin{equation}\label{eq:S_mmm}
			\mathcal S_{ab} = \frac{1}{\sqrt{2m-1}}\exp\left(-2\pi i\frac{2ab}{2m-1} \right)
		\end{equation} 
		and 
		\begin{align}
			\mathcal T_{ab} =\delta_{ab} e^{\frac{2\pi i }{12} } e^{2\pi i \frac{a^2}{2m-1}  },
		\end{align}
		where $a,b =0,1,\cdots,2m-2$ take integer numbers. Our theory also applies to two-component states that are not spin-singlets. In Appendix~Sec. IV, we give the modular matrices for the Halperin $(441)$ state, in which case the dimension of the modular matrices is $15$ and the $\bm a$ vectors in the coset $\Gamma^*_K/\Gamma_K$ form a two-dimensional lattice. 
		
		Moreover, our theory is not limited to the two-component case. For instance, for the three-component $(m,m,m,m-1)$ state with 
		$K=\left( \begin{array}
			{ccc}
			m&m-1&m-1\\m-1&m&m-1\\m-1&m-1&m
		\end{array} \right)$,
		the coset lattice is  $\Gamma_K^*/\Gamma_K=\{ \frac{1}{3m-2}(\bm e_1+\bm e_2+\bm e_3) |a=0,1,\cdots 3m-3\}$ [Fig. \ref{fig:Klattice}(b) depicts the case of $m=2$], and the corresponding modular matrices are
		\begin{equation}
			\mathcal{S}_{ab} = \frac{1}{3m-2} e^{ -6\pi i \frac{ab}{3m-2}}, \qquad
			\mathcal{T}_{aa} = e^{2\pi i(\frac{3a^2}{6m-4}-\frac{3}{24})}.
		\end{equation}

		\textit{Anyon condensation for spin-singlet states.---} 
		The central results in Eqs.~(\ref{eq:modularS1}) and (\ref{eq:modularT1}) can be further examined by several parallel methods. As an example, for the two-component spin-singlet Halperin $(m,m,m-1)$ state, we do not need to resort to the $K$ matrix when deriving the modular matrices. Instead  we rely on the CFT of the $\widehat{su(2)}_1$ WZW model with a $\widehat{u(1)}_{4m-2}$ boson which describes the $(m,m,m-1)$ state~\cite{MOORE1991362}. There are $8m-4$ primary fields in the CFT of the WZW model, whose topological spin and fusion rule are~\cite{yellowbook}
		\begin{align}
			&\qquad h_{(\lambda,a)} = \frac{\lambda(\lambda+2)}{12}+\frac{a^2}{8m-4},\\
			&(\lambda, a)\otimes(\mu, b) = \left( (\lambda+\mu)_{\mr{mod}~2},(a+b)_{\mr{mod}~4m-2}  \right) ,
		\end{align}
		with $\lambda,\mu=0,1$ and $a,b=0,1,\cdots,4m-3$. 
		Interestingly, there exists a special primary field [denoted as $J=(1,2m-1)$] with integer topological spin
		$ h_{(1,2m-1)}=\frac14+\frac{(2m-1)^2}{4(2m-1)}=\frac{m}{2}\in\mathbb{Z}$.
		In addition to the vacuum $\mathbf{1}$, the existence of such a bosonic field with integer spin points to an extended chiral algebra~\cite{yellowbook}, i.e.
		akin to that one can not distinguish $(\lambda,a)$ and $(\lambda,a)\otimes \mathbf{1}^l$ ($l\in\mathbb{N}$), one  can not distinguish $(\lambda,a)$ with $(\lambda,a)\otimes J^l$ ($l\in\mathbb{N}$). 
		After condensation of the bosonic fields, the $8m-4$ primary fields are mapped into $4m-2$ distinguishable primary fields $\{[(0,a)]_J|a=0,1,\cdots,4m-3 \}$, where $[\cdots]_J$ denotes a equivalent class under the fusion with $J$ field $[(0,a)]_J:(0,a)\sim(0,a)\otimes J$. 
		Moreover, we notice that the difference between the topological spin of $a$ and $a\otimes J$(in the sense of modulo 1) is
		\begin{equation}\nonumber
			h_{a\otimes J}-h_a = \frac14+\frac{(a+2m-1)}{8m-4}-\frac{a^2}{8m-4}=\frac{a}{2}.
		\end{equation}
		Thus only half anyons with even $a$ are deconfined~\cite{Slingerland_2009}, which form a set 
		$ \{ a\equiv[(0,2a)]_J|a=0,1,\cdots,2m-2 \} $
		with the topological spin $
		h_a = \frac{a^2}{2m-1}$ and fusion rule $\mathcal{N}_{ab}^{~~~c}$:
		$a\otimes b = (a+b)_{\mr{mod} \ 2m-1}$. 
		The braiding statistics of these deconfined anyons can be described by the $\mathcal{S}$ matrix 
		\begin{equation}\label{eq:anyoncond}
			\mathcal{S}_{ab} = \frac{1}{\sqrt{2m-1}}\exp\left(-2\pi i\frac{2ab}{2m-1} \right),
		\end{equation}
		where we have used the formula
		$	\mathcal S_{ab} = \frac{1}{\mathcal{D}}\sum_c \mathcal{N}_{a \bar b}^{~~~c} \exp\left[2\pi i (h_c-h_a-h_b)\right] \mr{d}_c $~\cite{Slingerland_2009}. Here $\mr{d}_c=1$ is the quantum dimension of Abelian anyon $c$, and $\mathcal{D}=\sqrt{2m-1}$ is the total quantum dimension. Thus, we conclude that, using the anyon condensation picture we can obtain exactly the same modular matrices as in Eq.~(\ref{eq:S_mmm}), without resorting to the explicit $K$ matrix. 
		See Appendix~Sec. III for more technical details.
		
		\textit{Numerical simulation for the Halperin $(221)$ state.---}
		So far our discussion are solely based on analytical derivations. Now we turn to numerical simulations in microscopic models to further validate our analytical results. A well-developed numerical recipe for the extraction of modular matrices is the ``minimal-entanglement state (MES)'' scheme~\cite{YiZhang2012,WZhu2013,WZhu2014,Liu2019}, which requires bipartition of the whole system in real space. In order to conveniently realize such bipartition, we consider two-component hardcore bosons in a checkerboard lattice model~\cite{Sun2011,Sheng2011,Neupert2011,Wang2011,Regnault2011}.
		The system's Hamiltonian takes the form of
		\begin{align}
			H=&\!\sum_{\sigma}\!\Big[-t\!\!\sum_{\langle\rr,\rr'\rangle}\!e^{i\phi_{\rr'\rr}}b_{\rr',\sigma}^{\dag}b_{\rr,\sigma}
			-\!\!\!\!\sum_{\langle\langle\rr,\rr'\rangle\rangle}\!\!\!t_{\rr,\rr'}'b_{\rr',\sigma}^{\dag}b_{\rr,\sigma}\nonumber\\
			&-t''\!\!\!\sum_{\langle\langle\langle\rr,\rr'\rangle\rangle\rangle}\!\!\!\! b_{\rr',\sigma}^{\dag}b_{\rr,\sigma}+H.c.\Big]+U\sum_{\rr}n_{\rr,\uparrow}n_{\rr,\downarrow}.
		\end{align}
		Here $b_{\rr,\sigma}^{\dag}$ creates a boson with pseudospin $\sigma=\uparrow,\downarrow$ at site $\rr$, and $\langle\ldots\rangle$,$\langle\langle\ldots\rangle\rangle$ and $\langle\langle\langle\ldots\rangle\rangle\rangle$ denote the nearest-neighbor, the next-nearest-neighbor, and the next-next-nearest-neighbor pairs of sites, respectively. 
		In what follows, we choose the parameters as $t'=0.3t$, $t''=-0.2t$, $\phi=\pi/4$~\cite{Sun2011}, and $U=4.0$.
		A robust Halperin $(221)$ state with three-fold nearly degenerate ground states ($|\xi_{i=1,2,3}\rangle$) can be stabilized in this system at the total filling
		$\nu=2/3$~\cite{TSZeng2017}. 
		
		\begin{figure}
			\centering
			\includegraphics[width=0.45\linewidth]{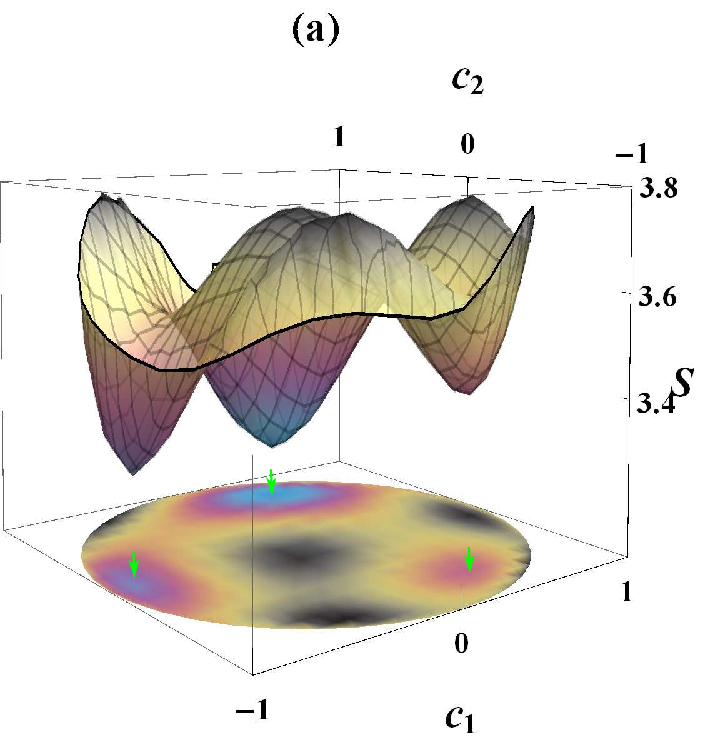} 
			\includegraphics[width=0.45\linewidth]{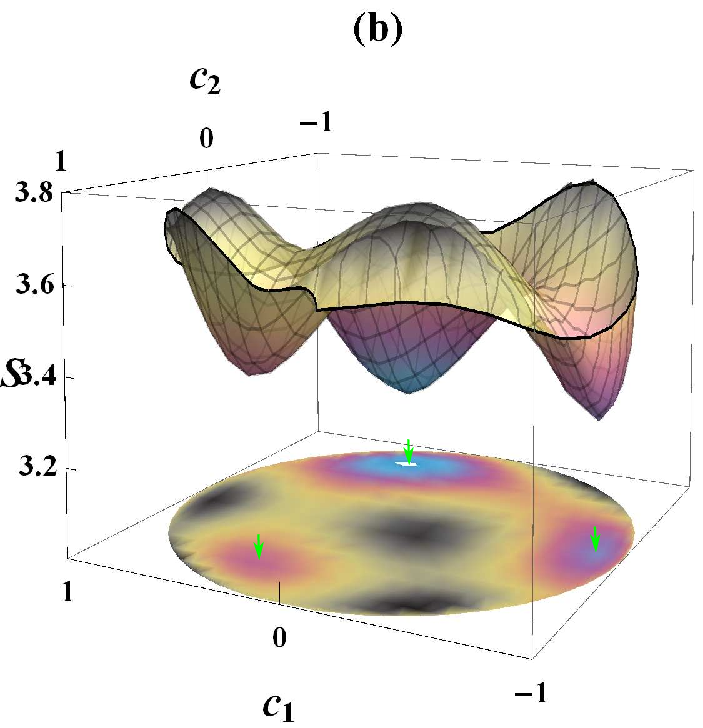}	
			\caption{\label{fig:MES}
				Surface and contour plots of the entanglement entropy $S$ of $|\Psi (c_1,c_2,\phi_2,\phi_3) \rangle $ versus $c_1$ and $c_2$ ($c_3 =\sqrt{	1 - c^2_1 - c^2_2}$). Here we fix $\phi_2$ and $\phi_3$ at their optimal values $\phi_2 = 0.5\pi, \phi_3 = 0.25\pi$. 
				The entanglement bipartition is along (a) the $x$ direction and (b) the $y$ direction. 
				The calculation is performed on a $3 \times 3$ checkerboard lattice at $\nu=2/3$.   
			}
		\end{figure}
		
		According to the MES scheme, we bipartite the lattice along $x$ and $y$ directions, respectively, and search for the corresponding MESs over all superpositions $|\Psi(c_1,c_2,\phi_2,\phi_3) \rangle = c_1|\xi_1 \rangle +c_2 e^{i\phi_2} |\xi_2 \rangle + c_3 e^{i\phi_3} |\xi_3 \rangle$ within the ground-state manifold, where $c_1, c_2, c_3, \phi_2, \phi_3$ are real parameters. 
		For each bipartition, by minimizing the entanglement entropy over the ranges of $c_i \in [0, 1]$ and $\phi_i \in [0,2\pi]$, we can identify three entropy valleys in the parameter space, each of which gives a global MES $|\Xi^{x,y}_{i=1,2,3}\rangle$ (Fig.~\ref{fig:MES}). 
		Then the overlap between the MESs along the two non-contractible bipartition directions gives the
		modular matrix $\mathcal S =\langle  \Xi^{y} | \Xi^{x} \rangle $~\cite{YiZhang2012,WZhu2013}. Our numerical calculation returns
		\begin{align}
			\mathcal S= \begin{pmatrix}
				0.587 & 0.572 & 0.572 \\
				0.572 & 0.581e^{-i0.67\pi} & 0.581e^{i0.67\pi} \\
				0.572 & 0.581e^{i0.67\pi} & 0.581e^{-i0.67\pi}
			\end{pmatrix}.
		\end{align} 
		This result is quite close to the theoretical prediction $ 	\mathcal S= \frac{1}{\sqrt{3}}\begin{pmatrix}
			1 & 1 & 1 \\
			1 & e^{-i2\pi/3} & e^{i2\pi/3} \\
			1 & e^{i2\pi/3} & e^{-i2\pi/3}
		\end{pmatrix}$ for the Halperin $(221)$ state,
		indicating the unit quasiparticle quantum dimension and the $Z_3$ mutual statistics
		between quasiparticles. Additionally, the statistics
		phases are in units of $2\pi/3$, which clearly signals the fractional
		charge $e/3$ of quasiparticles~\cite{ZLiu2019}.

		\textit{Summary and discussion.---}
		We have investigated the modular transformation and fractional statistics for Abelian multi-component fractional quantum Hall states. Crucially, we derive a universal
		formula for the modular matrices by utilizing the conformal field theory. We establish that the
		modular matrices can be fully formulated in terms of the $K$ matrix in the Chern-Simons field theory, which offers a first-principal evidence of the correspondence between the 2D CFT and (2+1)D topological field theory descriptions for multi-component FQH states. Moreover, we illustrate several examples to validate our theory. 
		One is the spin-singlet Halperin $(m,m,m-1)$ state obtained by the anyon condensation picture, and the other example is the Halperin $(221)$ state living on a microscopic lattice model, where the modular matrix are numerically computed via the minimal entangled states. 
		In both examples, the extracted modular matrices coincide with our theory.

		Apart from the field-theory methods, the modular matrices can also be directly calculated from the trial wave functions of FQH states. 
		By explicitly applying the modular $\mathcal S$ and $\mathcal T$ transformations on the trial wave function, we recover the CFT results of the modular matrices in Eqs.~(\ref{eq:modularS1}) and (\ref{eq:modularT1}) up to a particle-number-dependent phase factor (see Appendix~Sec. II for details).
		In this regard, modular matrices derived by the CFT can be understood via the trial wave function from a different perspective.

		Our work opens several directions for the future study. First, hopefully our analysis can be extended to other FQH states possessing $K$ matrices, such as those described by the affine Lie algebra conformal field theories~\cite{Ardonne2003} and the hierarchy FQH states \cite{Haldane1983}. It would be interesting to probe the relation between modular matrices and the $K$ matrix in those cases also. Second, investigation of modular matrices for lattice FQH states in Bloch bands with Chern number $C>1$ may be helpful to unveil the potential difference of these states from the usual $C$-component Halperin states~\cite{FCIHighC_Zhao,FCIHighC_Sterdyniak,FCIHighC_YangLe,ZhaoModelFCI,FCIHighC_Gunnar}. 
		Finally, one can study the modular matrices for FQH states beyond the $K$-matrix description, such as non-Abelian multi-component FQH states~\cite{Ardonne1999,Reijnders2002}.

		{\it Acknowledgments.---} We thank T. S. Zeng for collaborating on related projects.  
		LDH and WZ were supported by National Natural Science Foundation of China (No.~92165102, 11974288) and the foundation of Westlake University.  
		ZL was supported by National Natural Science Foundation of China No.~11974014.

		\bibliography{main}
		
		

		\onecolumngrid 
		\newpage
		
		\setcounter{subsection}{0}
		\setcounter{equation}{0}
		\setcounter{figure}{0}
		\renewcommand{\theequation}{S\arabic{equation}}
		\renewcommand{\thefigure}{S\arabic{figure}}
		
		\appendix
		In this supplementary material, we will show more details to support the discussion in the main text. In Sec.~I, we show the details of the derivation of modular matrices by the conformal field theory (CFT) that describes the edge. In Sec.~II, we calculate the modular matrices based on the trial wave function of the general Halperin state. In Sec.~III, we compute the modular matrices of the Halperin $(m,m,m-1)$ state, through an extended chiral algebra. In Sec.~IV, we present more examples that can be described by our theory.  
		
		\section{I. Conformal field theory method}
		\label{SMA}
		
		\subsection{1. Review of CFT for the Laughlin state}
		\label{SMA1}
		In this section, we first review the CFT which describes the edge of the single-component $\nu=1/(2q)$ Laughlin state. The simplest example in CFT is the free boson model $S = \frac{m}{4\pi}\int dx^2 \partial^\mu\phi(x)\partial_\mu\phi(x)$. We put this model on a torus (space-time manifold) and impose the following boundary condition
		\begin{equation}\nonumber
			\phi(z+k\omega_1+k'\omega_2)=\phi(z)+2\pi R(km+k'm')\quad k,k',m,m'\in\mathbb{Z},
		\end{equation}
		where $z$ is the coordinate in the torus, $\omega_1$ and $\omega_2$ are the periods of the torus (i.e. $z\sim z+\omega_1\sim z+\omega_2$). The doublet $(m,m')$ counts the winding number of $\phi$ when it goes around the torus in the periodic directions. Under this condition, $\phi$ is call the compactified boson on a circle with radius $R$. When $R^2$ is a rational number $R^2=\frac{2q}{p}$ with $p,q$ two co-prime positive numbers, this theory is called the rational CFT (RCFT). RCFT means that the Virasoro primary fields can be reorganized into finite number of extended blocks, and they are linearly covariant under the action of the modular group~\cite{yellowbook}.
		Here we set $p=1$, the partition function is
		\begin{equation}\label{boson_chi}
			Z(R=\sqrt{2q})=\frac{1}{|\eta(\tau)|^2}\sum_{e,m\in\mathbb{Z}}q^{\frac{(e+qm)^2}{4q}}\bar q^{\frac{(e-qm)^2}{4q}}
		\end{equation}
		with topological spins of primary fields
		$$ h_{e,m}=\frac{(e+qm)^2}{4q},\qquad\bar h_{e,m}=\frac{(e-qm)^2}{4q}. $$
		In Eq.~(\ref{boson_chi}), $\eta(\tau)=q^{1/24}\prod_{n=1}^{\infty}(1-q^n)$
		is the $\tau$-dependent Dedekind's $\eta$-function with $q=e^{2\pi \tau i}$, and $\tau=\frac{\omega_2}{\omega_1}$ is the parameter of the torus (see Fig.~1 in the main text). Those fields can be reorganized into a finite number of extended blocks
		\begin{equation}\begin{split}\nonumber
				Z(\sqrt{2q})&=\frac{1}{|\eta(\tau)|^2}\sum_{e,m\in\mathbb{Z}}q^{\frac{(e+qm)^2}{4q}}\bar q^{\frac{(e-qm)^2}{4q}}\\
				&=\frac{1}{|\eta(\tau)|^2}\sum_{n,m\in\mathbb{Z}}q^{\frac{n^2}{4q}}\bar q^{\frac{(n-2qm)^2}{4q}}\\
				&=\frac{1}{|\eta(\tau)|^2}\sum_{a=0}^{2q-1}\sum_{k,m\in\mathbb{Z}}q^{\frac{(a+2kq)^2}{4q}}\bar q^{\frac{(a+2(k-m)q)^2}{4q}}\\
				&=\frac{1}{|\eta(\tau)|^2}\sum_{a=0}^{2q-1}\sum_{k\in\mathbb{Z}}q^{\frac{(a+2qk)^2}{4q}}\sum_{\bar k\in\mathbb{Z}}\bar q^{\frac{(a+2q\bar k)^2}{4q}}\\
				&=\sum_{a=0}^{2q-1}|K_a^{(2q)}(\tau)|^2,
		\end{split}\end{equation}
		where the extended character of block $a$ is
		\begin{equation}
			K_a^{(N)}(\tau) = \frac{1}{\eta(\tau)}\sum_{n\in\mathbb{Z}}q^{\frac{(Nn+a)^2}{2N}}.
		\end{equation}
		We are interested in the transformation of $K_a^{(2q)}(\tau)$ under modular transformations.  The modular transformation $\tau\rightarrow \tau+1$ is easily seen to be
		\begin{equation}
			K_a^{(2q)}(\tau+1) = e^{2\pi i(\frac{a^2}{4q}-\frac{1}{24})}K_a^{(2q)}(\tau),
		\end{equation}
		where $\eta(\tau+1)=e^{\frac{2\pi i}{24}}\eta(\tau)$ and $e^{2\pi i\frac{(2qn+a)^2}{4q}}=e^{2\pi i\frac{a^2}{4q} }$.
		Another transformation $\tau\rightarrow -\frac{1}{\tau}$ is more involved:
		\begin{equation}\begin{split}
				K_a^{(2q)}(-1/\tau) & = \frac{1}{\sqrt{-i\tau}\eta(\tau)}\sum_n\exp\left[-\frac{ 2\pi i}{\tau} \frac{(2qn+a)^2}{4q} \right] \\
				& = \frac{1}{\sqrt{-i\tau}\eta(\tau)}\sum_n\exp\left[-\frac{ \pi i}{\tau}(2qn^2+2na+\frac{a^2}{2q}) \right] \\
				& = \frac{1}{\sqrt{2q}}\sum_k\exp\left[\pi i\tau\frac{(k-a/\tau)^2}{2q}-\frac{\pi i}{\tau}\frac{a^2}{2q} \right]\\
				& = \sum_k\frac{ e^{-2\pi i\frac{ak}{2q}} }{\sqrt {2q}} \exp\left(2\pi i\tau\frac{k^2}{4q}\right)\\
				&= \sum_{b=0}^{2q-1} \frac{ e^{-2\pi i\frac{ab}{2q}} }{\sqrt {2q}}  K_b^{(2q)}(\tau).
			\end{split}\label{eq:extendedS}\end{equation}
		Here we have used the Poisson resummation formula 
		$$ \sum_{n\in\mathbb{Z}}\exp(-\pi a n^2+bn)=\frac{1}{\sqrt{a}}\sum_{k\in\mathbb{Z}}\exp{\left[ -\frac{\pi}{a}(k+\frac{b}{2\pi i})^2\right] } $$
		in the third line. Therefore, the extended characters are linearly covariant under the modular transformations, and these results give the modular matrices
		\begin{align}
			\mathcal{T}_{ab} = e^{2\pi i(\frac{a^2}{4q}-\frac{1}{24})}\delta_{ab},\qquad
			\mathcal{S}_{ab} = \frac{ e^{-2\pi i\frac{ab}{2q}} }{\sqrt {2q}} .
		\end{align}
		The unitary of $\mathcal{S}$ does ensure the modular invariance of the partition function $Z(\sqrt{2q})=\sum_{a=0}^{2q-1}|K^{(2q)}_a(\tau)|^2$.
		In additional, the quantum dimensions of the fields are $\mr{d}_a=\frac{\mathcal{S}_{a0}}{\mathcal{S}_{00}}=1$.
		Using the Verlinde formula, we can derive the fusion rules of the extended fields
		\begin{equation}
			\mathcal{N}_{ab}^{~~~c} = \sum_{m} \frac{\mathcal{S}_{am}\mathcal{S}_{bm}\bar{\mathcal{S}}_{mc}}{\mathcal{S}_{0m}} = \delta^{\text{mod}~2q}_{a+b,c},
		\end{equation}
		where $\delta^{\text{mod}~2q}_{a,b}$ means $a=b\mod{2q}$.
		This theory is called the $\widehat{u(1)}_{2q}$ CFT~\cite{MOORE1991362,Dong2008,Wen_book,Wen1990,Wen1991,Wen1992}, where the level is square of the radius $R$. The $\widehat{u(1)}_{2q}$ CFT is the edge theory of bosonic Laughlin $\nu=1/(2q)$ state and is also the physical realization of the $\mathbb{Z}_{2q}^{(\frac12)}$ anyon model~\cite{Moore1989_ZN}.
		
		\subsection{2. Multi-component Abelian FQH states}\label{append:cft}
		\label{SMA2}
		In this section, we will extend the discussion of the CFT to general multi-component Abelian Halperin states. 
		One way to deal with the Halperin state is to generalize the $\widehat{u(1)}$ theory to $\widehat{u(1)^{\oplus_\kappa}}$~\cite{U1_level_k,Wen1992}, where $\kappa$ is 
		the number of components. The level of $\widehat{u(1)^{\oplus_\kappa}}$ is no longer a single number. Instead the level becomes a $\kappa\times \kappa$ positive-definite integer matrix $K$, which we will denote as 
		$\widehat{u(1)}_{\kappa,K}$.
		The simplest case is a diagonal $K$ with $K_{ii}$ are positive integers, thus this theory is just a simple direct sum of each $\widehat{u(1)}_{K_{ii}}$. Here we consider the bosonic states, thus the diagonal of $K$ is even. (Throughout this work we focus on bosonic states, and the extension to fermionic cases should be straightforward.) 
		
		In this context, the partition function of the multi-component state is written as 
		\begin{equation}\begin{split}
				Z(K) &= \sum_{\bm a \in \Gamma^*_K/\Gamma_K} |\chi_{\bm a}(\tau)|^2 \\
				\chi_{\bm a}(\tau) &= \frac{1}{\eta(\tau)^\kappa}\sum_{\bm n\in\Gamma_K}q^{\frac12 (\bm n+\bm a)\cdot (\bm n+\bm a)}.
		\end{split}\end{equation}
		Here we use an intuitive way to define the topological sectors. That is, we define the so-called a $K$-lattice $ \Gamma_K =\{\bm n=\sum_I n_I \bm e_I|n_I\in\mathbb{Z}, \bm e_I\cdot\bm e_J=K_{IJ}\}$. And its dual lattice $\Gamma_K^*$ is spanned by the dual basis $\bm e^*_I$ satisfying 	the relation $\bm e_I\cdot \bm e^*_J=\delta_{IJ}$. $\Gamma^*_K/\Gamma_K$ is the coset (see Fig.~1 in the main text). The transformation $\tau\rightarrow \tau+1$ can be easily evaluated with the assumption that $K_{ii}$ is even:
		\begin{equation}
			\chi_{\bm a}(\tau+1) = e^{2\pi i(\frac12 \bm a \cdot \bm a-\frac{\kappa}{24})}\chi_{\bm a}(\tau).
		\end{equation}
		So we reach 
		\begin{align}
			\mathcal T =e^{2\pi i(\frac12 \bm a \cdot \bm a-\frac{\kappa}{24})}.
		\end{align}

		In order to consider the $\mathcal S$ transformation $\tau\rightarrow -\frac{1}{\tau}$, we utilize the generalized Poisson's resummation formula
		\begin{equation}\nonumber
			\sum_{\bm q\in\Gamma_K}e^{-\pi a \bm q^2 + \bm q \cdot \bm b} = \frac{1}{\sqrt{|\det K|}a^{\frac{\kappa}{2}}}\sum_{\bm p\in \Gamma^*_K}e^{-\frac{\pi}{a}(\bm p+\frac{\bm b}{2\pi i})^2},
		\end{equation}
		and apply it to the character. Then we have 
		\begin{equation}\label{A8}\begin{split}
				\chi_{\bm a}(-1/\tau)&= \frac{1}{(-i\tau)^{\kappa/2}\eta(\tau)^\kappa}\sum_{\bm n\in\Gamma_K}e^{-\frac{\pi i}{\tau}\left(\bm n\cdot \bm n+2\bm n\cdot \bm a+\bm a\cdot\bm a \right) }\\
				&=\frac{1}{\sqrt{|\det K|}}\frac{1}{\eta(\tau)^\kappa}\sum_{\bm m\in\Gamma_K^*}e^{\pi i \tau \bm m\cdot \bm m-2\pi i \bm m\cdot \bm a}.
		\end{split}\end{equation}
		The finial step is to rewrite the summation over $\Gamma_K^*$, by noticing the vector in $\Gamma_K^*$ can be written as 
		$$\bm m=\sum_I m_I \bm e_I^*=\sum_I\left( \sum_J n_J K_{JI} + b_I \right)e_I^* =\bm n+\bm b $$ 
		with $\bm n=\sum_I n_I \bm e_I\in \Gamma_K, \bm b=\sum_I b_I \bm e^*_I\in \Gamma_K^*/\Gamma_K$ and the linear combination of dual basis $\sum_J K_{IJ}\bm e_J^*$ is exactly the basis of $\Gamma_K$: 
		$ \bm e_I \cdot \bm e_J=\sum_N K_{IN}\bm e_N^*\cdot\bm e_J=K_{IJ} $. Thus, we can rewrite Eq.~(\ref{A8}) as 
		\begin{equation}\begin{split}
				\chi_{\bm a}(-1/\tau)&=\frac{1}{\sqrt{|\det K|}}\frac{1}{\eta(\tau)^\kappa}\sum_{\bm m\in\Gamma_K^*}e^{\pi i \tau \bm m\cdot \bm m-2\pi i \bm m\cdot \bm a}\\
				&=\sum_{\bm b\in \Gamma_K^*/\Gamma_K}\frac{e^{-2\pi i \bm a\cdot\bm b}}{\sqrt{|\det K|}} \chi_{\bm b}(\tau).
		\end{split}\end{equation}
		The prefactor is just the modular $\mathcal{S}$ matrix
		\begin{equation}\label{eq:a_S_abelian}
			\mathcal{S} = \frac{e^{-2\pi i \bm a\cdot\bm b}}{\sqrt{|\det K|}}.
		\end{equation}
		This is the result shown in the main text.
		
		At last, as an example, we consider the $(m,m,m-1)$ state, for which the $K$ matrix is 
		$\begin{pmatrix}
			m & m-1\\ m-1 & m
		\end{pmatrix}$
		and the coset $\Gamma^*_K/\Gamma_K$ contains $|\det K|=2m-1$ independent vectors $\bm a=\{ \frac{1}{2m-1}(a\bm e_1+a\bm e_2)|a=0,1,\cdots,2m-2 \}$ ($\bm b$ is defined similarly). We can write the $\mathcal T$ matrix as
		\begin{align}
			\mathcal T_{ab} =\delta_{ab} e^{\frac{2\pi i }{12} } e^{2\pi i \frac{a^2}{(2m-1)^2}  }
		\end{align}
		and the
		$\mathcal{S}$ matrix as
		\begin{equation}\label{res_WF}
			\mathcal S_{ab} = \frac{1}{\sqrt{2m-1}}\exp\left(-2\pi i\frac{2ab}{2m-1} \right).
		\end{equation}
		For more examples please see Sec.~\ref{SMD}. 
		

		\section{II. Trial wave function method}\label{append:trial_wf}
		
		In this section, we introduce another method for deriving the modular matrices. This method is based on the $K$-matrix related trial wave function. This is method is independent of the CFT, which can be viewed as a complementary method to the CFT. Part of results have overlaps with an unpublished work~\cite{Wen2012}. 
		
		\subsection{1. Gauge transformation}
		Let us consider the Hamiltonian of a charged electron on the torus spanned by $\vec L_1 = L \vec e_x$ and $\vec L_2 = L\vec\tau$ with a uniform perpendicular magnetic field \cite{Hu2021}
		\begin{align}\label{H-tau-s}
			H_0(\mathbf{A},\tau) &= \frac12 g^{ab}(\tau)D_a(\mathbf{A}) D_b(\mathbf{A}), 
		\end{align}
		where  $D_a(\mathbf{A})=-i\hbar\partial/\partial X^a+|e|A_a$ and $\mathbf{A}=(-\tau_2 L^2 BX^2,0)$ are the covariant derivative and vector potential, respectively. $g(\tau)$ is the $\tau$-dependent metric
		\begin{align}
			g(\tau) &=\frac{1}{S\tau_2}
			\left(
			\begin{array}
				{cc}
				|\tau|^2	&	-\tau_1\\
				-\tau_1		&	1
			\end{array}
			\right),
		\end{align}
		where $S=|\vec L_1\times \vec L_2|=\tau_2L^2$ is the area of the torus which is invariant under any modular transformation. The ground state is the lowest Landau level (LLL)
		\begin{equation}\label{Landau-level}
			\Psi_m(X^1,X^2|\tau) = \frac{1}{\sqrt{\pi^{1/2}L\ell}}\mathrm{e}^{i\pi N_{\phi}\tau [X^2]^2}
			\theta_{\frac{m}{N_\phi}}(N_\phi z/L|N_\phi \tau)
		\end{equation}
		with $m=0,1,\cdots,N_\phi-1$, where $N_\phi=\frac{\tau_2 L^2}{2\pi\ell^2}$ is the total flux through the torus, $\ell=\sqrt{\hbar/|e|B}$ is the magnetic length,
		$z=x+iy=L(X^1+\tau X^2)$ is the complex coordinate of electron, and
		$\theta_{m}(z|\tau)$ is the theta function defined as
		\begin{equation}
			\theta_{\alpha}(z|\tau)=\sum_{n\in Z} \exp{\left( i\pi \tau (n+\alpha)^2 +i2\pi(n+\alpha)z \right) }.
		\end{equation}
		
		The modular $S$ transformation makes a $\pi/2$ rotation in $(X^1,X^2)$ space
		\begin{equation}\label{d-transformation}
			\left.
			\left(
			\begin{array}
				{c}
				X'^1	\\
				X'^2
			\end{array}
			\right)
			=
			\left(
			\begin{array}
				{cc}
				0	&	-1\\
				1		&	0
			\end{array}
			\right)
			\left(
			\begin{array}
				{c}
				X^1	\\
				X^2
			\end{array}
			\right)
			\right.
		\end{equation}
		and transforms $\tau$ to $-1/\tau$. Since the exchange of the two sides of the parallelogram (torus), the torus is spanned by $\vec L'_1 = |\tau|L \vec e_x, ~~\vec L'_2 = L\frac{(-\tau_1,\tau_2)}{|\tau|}$ and the complex coordinate should be expressed as $z'= |\tau|L(X'^1-\frac{1}{\tau}X'^2)=\frac{|\tau|}{\tau}z=\frac{\tau^*}{|\tau|}z$. 
		After the $S$ transformation, the single-particle Hamiltonian becomes
		\begin{align}\label{H-tau1-s}
			H_0(\mathbf{A}',-1/\tau) &= \frac12 g^{ab}(-1/\tau)D'_a(\mathbf{A}') D'_b(\mathbf{A}')
		\end{align}
		with
		\begin{align}
			g(-1/\tau) &=\frac{1}{S\tau_2}
			\left(
			\begin{array}
				{cc}
				1	&	\tau_1\\
				\tau_1		&	|\tau|^2
			\end{array}
			\right),
		\end{align}
		and the LLL wave function becomes 
		\begin{eqnarray}\label{Landau-level-1}
			\Psi_m(X'^1,X'^2|-1/\tau) = \frac{1}{\sqrt{\pi^{1/2}L\ell}}\mathrm{e}^{i\pi 	N_{\phi}\frac{1}{-\tau} [X'^2]^2}\nonumber\\
			\times\theta_{\frac{m}{N_\phi}}(N_\phi z'/L|N_\phi (-1/\tau)),
		\end{eqnarray}
		where $D'_a(\mathbf{A}')=-i\hbar\partial/\partial X'^a+|e|A'_a$ and $\mathbf{A}'=(-\tau_2 L^2 BX'^2,0)$ are the covariant derivative and vector potential in $(X'^1,X'^2)$ space, respectively. 
		To compare Eq.~(\ref{Landau-level-1}) with Eq.~(\ref{Landau-level}), we need to write them in the same coordinate frame.
		Therefore, $H_0(\mathbf{A}',\tau+1)$ should be rewritten in $(X^1,X^2)$ by using the relation
		\begin{equation}\label{g-tau+1}
			g(-1/\tau)=
			M_S^\dagger g(\tau) M_S
		\end{equation}
		with
		\begin{equation}
			M_S=\left(
			\begin{array}
				{cc}
				0	&	1\\
				-1		&	0
			\end{array}
			\right).
		\end{equation}
		Thus, Eq.~(\ref{H-tau1-s}) can be written as
		\begin{align}
			H_0(\mathbf{A}',-1/\tau) &= \frac12 \bm (M_S \bm D'(\mathbf{A}'))_a g^{ab}(\tau) (M_S \bm D'(\mathbf{A}'))_b.
		\end{align}
		After denoting
		\begin{equation}\label{d-transformation}
			\left.
			\left(
			\begin{array}
				{c}
				D_1(\tilde{\mathbf{A}})	\\
				D_2(\tilde{\mathbf{A}})	
			\end{array}
			\right)
			=M_S \bm D'(\mathbf{A}') = M_S
			\left(
			\begin{array}
				{c}
				D'_1(\mathbf{A}')	\\
				D'_2(\mathbf{A}')	
			\end{array}
			\right)
			\right.
		\end{equation}
		with $\tilde{\mathbf{A}}=(0,\tau_2 L^2 B X'^2)=(0,\tau_2 L^2 B X^1)$, 
		we find
		\begin{equation}\label{H-tau1-s-re}
			H_0(\mathbf{A}',-1/\tau) =H_0(\tilde{\mathbf{A}},\tau)= \frac12 g^{ab}(\tau)D_a(\tilde{\mathbf{A}}) D_b(\tilde{\mathbf{A}})  
		\end{equation}
		with $D_a(\tilde{\mathbf{A}})=-i\hbar\partial/\partial X^a+|e|\tilde{A}_a$. Since $D_a(\mathbf{A})$ and $D_a(\tilde{\mathbf{A}})$ can be related by a gauge transformation $\hat{\mathcal{U}}=\mathrm{e}^{-2\pi iN_{\phi}X^1X^2}$, i.e.,
		\begin{equation}\label{D-gauge}
			D_a(\mathbf{A})=\hat{\mathcal{U}}^\dagger D_a(\tilde{\mathbf{A}})\hat{\mathcal{U}},
		\end{equation}
		we get
		\begin{equation}\label{H-gauge}
			H_0(\mathbf{A},\tau)=\hat{\mathcal{U}}^\dagger H_0(\mathbf{A}',-1/\tau)\hat{\mathcal{U}}.
		\end{equation}
		Now we can see that the LLL wave function after this transformation is
		\begin{equation}
			\begin{split}\nonumber
				&\hat{\mathcal{U}}\Psi_m(X'^1,X'^2|-1/\tau)\\
				& = \frac{\mr{e}^{i\pi N_\phi\frac{1}{-\tau}[X'^2]^2-2\pi i N_\phi X^1 X^2}}{\sqrt \pi L\ell}\theta_{\frac{m}{N_\phi}}\left(N_\phi z'\bigg|-\frac{N_\phi}{\tau}\right) \\
				& = \frac{\mr{e}^{-i\pi N_\phi \frac{z^2}{\tau L^2}-i\pi\tau N_\phi [X^2]^2}}{\sqrt \pi L\ell}\theta_{\frac{m}{N_\phi}}\left(N_\phi\frac{|\tau|z}{\tau}\bigg|-\frac{N_\phi}{\tau} \right)
			\end{split}
		\end{equation}
		The Poisson's resummation formula gives us
		\begin{equation}\nonumber
			\begin{split}
				&\theta_{\frac{m}{N_\phi}}\left(N_\phi\frac{|\tau|z}{\tau}\bigg|-\frac{N_\phi}{\tau} \right)\\
				&= \sqrt{\frac{-i\tau}{N_\phi}}\sum_{n=0}^{N_\phi-1}
				\theta_{\frac{n}{N_\phi}}\left(N_\phi\frac{ z}{L}\bigg|N_\phi\tau \right) \mr e^{ -2\pi i\frac{mn}{N_\phi}+i\pi N_\phi \frac{z^2}{\tau L^2} }.
			\end{split}
		\end{equation}
		This suggests us to redefine the wave function $\Psi(X^1,X^2|\tau)$ as $\Psi(X^1,X^2|\tau)/\eta(\tau)$. The newly defined wave function satisfies the modular covariance
		\begin{equation}
			\hat{\mathcal{U}}\frac{\Psi_m(X'^1,X'^2|-1/\tau)}{\eta(-1/\tau)} = \sum_{n=0}^{N_\phi-1}\mathcal S_{mn}\frac{\Psi_n(X^1,X^2|\tau)}{\eta(\tau)},
		\end{equation}
		where $\mathcal S_{mn} = \frac{1}{\sqrt{N_\phi}}\exp(-2\pi i\frac{mn}{N_\phi})$. One may ask why the single-orbital wave function under $S$ transformation can not be directly related by $ \mathcal{\hat{U}} \Psi_m(-1/\tau)=\Psi_m(\tau)$. The reason is degeneracy. We denote the space of degenerate LLL wave functions by $\mathcal M(\tau)$, for which 
		$$ \mathcal M(\tau) = \mathcal{\hat{U}} \mathcal M(-1/\tau) = \mathcal{\hat{U}} \mathcal{\hat{S}} \mathcal M(\tau). $$
		Thus, the action of $S$ transformation will induce a unitary transformation on single state:
		$$ \mathcal{\hat{U}} \mathcal{\hat{S}} \Psi_m(\tau) = \mathcal{\hat{U}} \Psi_m(-1/\tau) = \mathcal{S}_{mn} \Psi_n(\tau), $$
		where $\mathcal S_{mn}$ on the right hand of above equation is the representation of modular $S$ transformation in space $\mathcal M(\tau)$:
		$$ \langle \Psi_n;\tau|\mathcal{\hat S}|\Psi_m;\tau\rangle = \mathcal S_{mn} .$$
		Here the gauge transformation are implicitly included in the Dirac notation since we can not compare wave functions in different gauges:
		\begin{equation}
			\begin{split}
				&\langle \Psi_n;\tau|\mathcal{\hat S}|\Psi_m;\tau\rangle \\
				= &\int dX^1dX^2 \Psi^*(X^1,X^2|\tau)\mathcal{\hat{U}} \Psi(X'^1,X'^2|-1/\tau) .
			\end{split}
		\end{equation}

		

		\subsection{2. FQH wave function and modular $\mathcal{S}$ matrix}
		On the torus geometry, the general Halperin state can be expressed in terms of the theta-function~\cite{Hansson2008,Hansson2009,Hansson2014,Wen1995}:
		\begin{eqnarray}\label{Abelian-Wave_append}
			\Psi^{\bm{ a}}\left(\left\lbrace z_i^I \right\rbrace| \tau\right) &=& \mathcal{N}(\tau)
			f^{\bm a}_c(\left\lbrace Z^I \right\rbrace ) f_r( \left\lbrace z_i^I \right\rbrace ) 
			\times\exp{\left\lbrace i\pi \tau N_{\phi} \sum_{I,i}\left( \frac{y_i^I}{L\tau_2}\right)^2\right\rbrace},
		\end{eqnarray}
		where $I$ is the index of component, $\bm{a}\in \Gamma_K^*/\Gamma_K$ is the vector labeling degenerate states, $z^I_i=L(X^{I1}_i+\tau X^{I2}_i)$ is the coordinate of the $i$th particle in the $I$th component, and $Z^I=\sum_i z^I_i$ is the center-of-mass coordinate of the $I$th component. 
		$f_r$ and $f_c$ are the relative part and center of mass part of the wave function:
		\begin{eqnarray}
			f_r\left( \left\lbrace z_i^I \right\rbrace| \tau \right) &=& \left\lbrace \prod_{I<J} \prod_{i,j} \eta^{-K_{IJ}}(\tau) \theta^{K_{IJ}}_{11}(z_i^I/L-z_j^J/L|\tau) \right\rbrace\nonumber\\
			&\times&\left\lbrace \prod_{I} \prod_{i<j} \eta^{-K_{II}}(\tau) \theta^{K_{II}}_{11}(z_i^I/L-z_j^I/L|\tau) \right\rbrace,\nonumber\\
			f^{\bm a}_c\left(\left\lbrace Z^I \right\rbrace | \tau\right) &=& \eta^{-\kappa}(\tau) f^{(\bm{a})}(\bm{Z}/L|\tau),
		\end{eqnarray}
		where $K_{IJ}$ is the underlying $K$ matrix with dimension $\dim(K)=\kappa$ and diagonal elements $\bm{\kappa}=(K_{11},K_{22},\cdots,K_{\kappa\kappa})^T$,
		\begin{equation}
			\begin{split}
				\theta_{11}(z|\tau)&=\sum_{n\in \mathbb{Z}} \exp{\left\lbrace i\pi\tau \left(n+\frac12\right)^2 +i2\pi \left(n+\frac12\right)\left(z+\frac12\right) \right\rbrace },\\
				f^{(\bm{a})}(\bm{Z}|\tau) &= \sum_{\bm{n}\in\Gamma_K} 
				\exp{ \left\lbrace i\pi\tau(\bm{n}+\bm{a})^2 +2\pi i(\bm{n}+\bm{a})\cdot\bm{Z} \right\rbrace  },
			\end{split}
		\end{equation}
		and $\bm Z=\sum_I Z^I \bm e_I$ with $\bm e_I$ the basis of $\Gamma_K$ (i.e. $\bm e_I\cdot \bm e_J = K_{IJ}$).
		The normalization factor is 
		\begin{equation}\label{Normalization}
			\mathcal{N}(\tau) = N_0 \left[ \sqrt{\tau_2} \eta(\tau)^2 \right]^{\frac{1}{2} \sum_I \kappa_I N^I}, 
		\end{equation}
		where $\eta(\tau)=q^{1/24}\prod_{n=1}^{\infty}(1-q^n)|_{q=e^{i2\pi\tau}}$
		is the $\tau$-dependent Dedekind's $\eta$-function, and $N_0$ is an area-dependent constant. There are more detailed discussions of the FQH wave function on the torus in Ref.~\cite{Hansson2008}. 
		
		Now, we can extract the modular $\mathcal S$ matrix by considering
		\begin{equation}\label{def_s}
			\begin{split}
				\mathcal{S}_{\bm b\bm a}&=\langle \Psi^{\bm b};\tau|\mathcal{\hat S}|\Psi^{\bm a};\tau\rangle \\&= \int\prod_{I,i}d z_i^I
				\Psi^{\bm{ b}*}\left(\left\lbrace z_i^I \right\rbrace| \tau\right)\mathcal{\hat U}_g \Psi^{\bm{ a}}\left(\left\lbrace z_i^{\prime I} \right\rbrace\bigg| \frac{1}{\tau}\right),
			\end{split}
		\end{equation}
		where $\mathcal{\hat U}_g$ the gauge transformation
		\begin{equation}\label{many-body-gauge}
			\hat{\mathcal{U}}_g = \exp{\left\lbrace -2\pi i N_{\phi} \sum_{I,i} X^{1,I}_i X^{2,I}_i \right\rbrace}.
		\end{equation}
		derived above, and $z'=|\tau|z/\tau$ is the coordinate after modular $S$ transformation. 
		Before the tedious derivation, we list some useful relations (here we assume the total flux $N_{\phi}$ through the torus is even):
		\begin{eqnarray}\label{Modular-Properties}
			\eta(-1/\tau) &=& \sqrt{-i\tau}\eta(\tau),\nonumber\\
			\theta_{11}(z/\tau|-1/\tau)    &=& -\sqrt{-i\tau}e^{\frac{i\pi z^2}{\tau}}\theta_{11}(z|\tau),\nonumber\\
			f^{(\bm{a})}(\bm{Z}/\tau|-1/\tau)  &=&   (-i\tau)^{\kappa/2}e^{i\pi \bm{Z}^2/\tau}
			\sum_{\bm b\in \Gamma^*_K/\Gamma_K}\frac{e^{-2\pi i\bm a \cdot\bm b}}{\sqrt{|\det K|}}f^{(\bm{b})}(\bm{Z}|\tau) .\nonumber
		\end{eqnarray}
		
		Next we deal with the transformation term by term.
		
		{\it 1. The Derivation of $\mathcal{N}(-1/\tau)$}\\
		\begin{equation}\label{part1}
			\begin{split}
				\mathcal{N}(-1/\tau) &= N_0\left[ \frac{\sqrt\tau_2}{|\tau|}\eta\left(\frac{1}{-\tau}\right)^2 \right]^{\frac12 \sum_I \kappa_I N^I} 
				= \left( \frac{-i\tau}{|\tau|} \right)^{\frac12\sum_I \kappa_I N^I}\mathcal{N}(\tau),
			\end{split}
		\end{equation}
		where we have used the first relation in Eq.~(\ref{Modular-Properties}).
		
		{\it 2. The center of mass part $f_c(\lbrace Z^I\rbrace|-1/\tau)$ }
		\begin{equation}
			\begin{split}
				f^{\bm a}_c&\left( \lbrace Z^{\prime I}\rbrace|-1/\tau \right) = \eta^{-\kappa}(-1/\tau)f^{(\bm a)}\left( \frac{\bm Z}{\tau L}\bigg|\frac{1}{-\tau} \right) \\
				& = \mr{e}^{i\pi\frac{\bm Z^2}{\tau L^2}} \sum_{\bm b\in\Gamma_K^*/\Gamma_K}\frac{e^{-2\pi i\bm a \cdot\bm b}}{\sqrt{|\det K|}}f^{\bm b}_c(\{Z^I\}|\tau),
			\end{split}
		\end{equation}
		where we have used the first and third relations in Eq.~(\ref{Modular-Properties}).
		
		{\it 3. The relative part $f_r(\{z_i^I\}|\tau)$}
		\begin{equation}
			\begin{split}
				&f_r(\{z_i^{\prime I}\}|-1/\tau)\\
				&=\mr{e}^{\frac{i\pi}{\tau L^2}\left( \sum_{I<J}\sum_{i\in I,j\in J} K_{IJ}(z^{IJ}_{ij})^2+\sum_{I}\sum_{i<j}K_{II}(z^{II}_{ij})^2\right)} \\
				&\times(-1)^{\sum_{I<J}\sum_{i\in Ij\in J} K_{IJ} + \sum_{I}\sum_{i<j}K_{II}}f_r(\{z_i^{I}\}|\tau),
			\end{split}
		\end{equation}
		where $z^{IJ}_{ij}=z^I_i-z^J_j$ and the first two relations in Eq.~(\ref{Modular-Properties}) have been used. The phase factor can be simplified. The constant phase is
		\begin{equation}
			\begin{split}
				&\sum_{I<J}\sum_{i\in Ij\in J} K_{IJ} + \sum_{I}\sum_{i<j}K_{II}\\
				& = \sum_{I<J}N^I K_{IJ} N^J + \sum_{I}\frac12 (N^I-1)N^I K_{II}\\
				& = \frac12 \left( \bm N^2-\sum_I \kappa_I N^I \right),
			\end{split}
		\end{equation}
		where $\bm N=(N_1,N_2,\cdots)$ is the particle numbers. The coordinate-dependent phase needs a careful treatment, since they should be fully canceled when putting
		all parts together:
		\begin{equation}
			\begin{split}
				&\frac{i\pi}{\tau L^2}\left( \sum_{I<J}\sum_{i\in I,j\in J} K_{IJ}(z^{IJ}_{ij})^2+\sum_{I}\sum_{i<j}K_{II}(z^{II}_{ij})^2\right)\\
				=&\frac{i\pi}{2\tau L^2}\sum_{I,J}\sum_{i\in I,j\in J} K_{IJ}(z^{IJ}_{ij})^2 \\
				=&\frac{i\pi}{\tau L^2}\sum_{I,J}\sum_{i\in I,j\in J}\left( z_i^IK_{IJ}z^I_i-z_i^IK_{IJ}z^J_j \right) \\
				=&-\frac{i\pi}{\tau L^2}\bm Z^2+\frac{i\pi}{\tau L^2}\sum_{I,i\in I}z_i^Iz_i^I\sum_{J}K_{IJ}N^J \\ 
				=&-i\pi\frac{\bm Z^2}{\tau L^2}+\frac{i\pi N_\phi}{\tau L^2}\sum_{I,i}(z_i^I)^2.
			\end{split}
		\end{equation}
		
		{\it 4. Exponential and gauge transformation} \\
		Since $y^I_i/(\tau_2 L) = X_i^{I,2}$, we have
		\begin{equation}\label{part4}
			\begin{split}
				&\mathcal{\hat U}_g \exp{\left\lbrace i\pi\frac{1}{-\tau} N_{\phi} \sum_{I,i}\left( X_i^{\prime I,2} \right)^2\right\rbrace} \\
				=&  \mathcal{\hat U}_g \exp{\left\lbrace -\frac{i\pi N_{\phi}}{\tau}  \sum_{I,i}\left(X_i^{I,1} \right)^2 \right\rbrace}\\
				=&  \exp{\left\lbrace -\frac{i\pi N_{\phi}}{\tau}\sum_{I,i}(z_i^I)^2 \right\rbrace}
				\exp{\left\lbrace i\pi\tau N_{\phi} \sum_{I,i}\left( X_i^{I,2} \right)^2\right\rbrace}.
			\end{split}
		\end{equation}
		Substituting Eq.~(\ref{Abelian-Wave_append}) and Eqs.~(\ref{part1})-(\ref{part4}) into Eq.~(\ref{def_s}), we finally have
		\begin{equation}\label{S-matrix}
			\langle \Psi^{\bm{b}};\tau|\mathcal{S}|\Psi^{\bm{a}};\tau\rangle = 
			\left( \frac{i\tau}{|\tau|} \right)^{\frac12 \sum_I \kappa_I N^I}
			\left( -1 \right)^{\frac12 \bm{N}^2} \frac{e^{-2\pi i\bm a\cdot \bm b}}{\sqrt{|\det K|}},
		\end{equation}
		where $\bm N=\sum_I N^I \bm e_I$ is a vector enclosing the particle number in each component $N^I$. We expect that  the intrinsic properties such as braiding statistics should be independent on the particle number $\bm N$. 
		If we neglect the particle number dependent $U(1)$ factor~\cite{Wen2016}, we have derived the modular $\mathcal{S}$ matrix from the trial wave function:
		\begin{equation}\label{WF_S}
			\mathcal{\bm S} = \frac{e^{-2\pi i\bm a \cdot \bm b}}{\sqrt{|\det K|}}.
		\end{equation}
		The above result is consistent with Eq.~(\ref{eq:a_S_abelian}) and Eq.~(4) in the main text.

		\subsection{3. Modular $\mathcal T$ transformation}
		Let us firstly consider how the LLL wave functions evolve under the Dehn-twist transformation $\tau\rightarrow\tau+1$. After the Dehn twist, the coordinate
		$z=L(X^1+\tau X^2)$ can be rewritten as $z=L(X'^1+(\tau+1) X'^2)$. Thus we express the single-particle Hamiltonian in terms of $(X'^1,X'^2)$ as
		\begin{align}\label{H-tau1-s}
			H_0(\mathbf{A}',\tau+1) &= \frac12 g^{ab}(\tau+1)D'_a(\mathbf{A}') D'_b(\mathbf{A}')
		\end{align}
		with
		\begin{align}
			g(\tau+1) &=\frac{1}{L^2\tau_2^2}
			\left(
			\begin{array}
				{cc}
				|\tau+1|^2	&	-\tau_1-1\\
				-\tau_1-1		&	1
			\end{array}
			\right),
		\end{align}
		and the LLL wave function takes the form of 
		\begin{eqnarray}\label{Landau-level-1}
			\Psi_m(X'^1,X'^2|\tau+1) = \frac{1}{\sqrt{\pi^{1/2}L\ell}}\mathrm{e}^{i\pi N_{\phi}(\tau+1) [X'^2]^2}\nonumber\\
			\times\theta_{\frac{m}{N_\phi}}(N_\phi z/L|N_\phi (\tau+1)),
		\end{eqnarray}
		where $D'_a(\mathbf{A}')=-i\hbar\partial/\partial X'^a+|e|A'_a$ and $\mathbf{A}'=(-\tau_2 L^2 BX'^2,0)$ are the covariant derivative and vector potential in $(X'^1,X'^2)$, respectively. 
		To compare Eq.~(\ref{Landau-level-1}) with Eq.~(\ref{Landau-level}), we need to write them in the same coordinate frame.
		Therefore, $H_0(\mathbf{A}',\tau+1)$ should be rewritten in $(X^1,X^2)$. By using relations
		\begin{equation}\label{g-tau+1}
			\left.
			g(\tau+1)=\frac{1}{L^2\tau_2^2}
			\left(
			\begin{array}
				{cc}
				1	&	-1\\
				0		&	1
			\end{array}
			\right)
			g(\tau)
			\left(
			\begin{array}
				{cc}
				1	&	0\\
				-1		&	1
			\end{array}
			\right)
			\right.
		\end{equation}
		and 
		\begin{equation}\label{d-transformation}
			\left.
			\left(
			\begin{array}
				{c}
				D_1(\tilde{\mathbf{A}})	\\
				D_2(\tilde{\mathbf{A}})	
			\end{array}
			\right)
			=
			\left(
			\begin{array}
				{cc}
				1	&	0\\
				-1		&	1
			\end{array}
			\right)
			\left(
			\begin{array}
				{c}
				D'_1(\mathbf{A}')	\\
				D'_2(\mathbf{A}')	
			\end{array}
			\right)
			\right.
		\end{equation}
		with $\tilde{\mathbf{A}}=(-\tau_2 L^2 B X'^2,\tau_2 L^2 B X'^2)=(-\tau_2 L^2 B X^2,\tau_2 L^2 B X^2)$, 
		we find
		\begin{equation}\label{H-tau1-s-re}
			H_0(\mathbf{A}',\tau+1) =H_0(\tilde{\mathbf{A}},\tau)= \frac12 g^{ab}(\tau)D_a(\tilde{\mathbf{A}}) D_b(\tilde{\mathbf{A}})  
		\end{equation}
		with $D_a(\tilde{\mathbf{A}})=-i\hbar\partial/\partial X^a+|e|\tilde{A}_a$. Because $D_a(\mathbf{A})$ and $D_a(\tilde{\mathbf{A}})$ can be related by a gauge transformation $\hat{\mathcal{U}}=\mathrm{e}^{-i\pi N_{\phi}[X^2]^2}$, i.e.,
		\begin{equation}\label{D-gauge}
			D_a(\mathbf{A})=\hat{\mathcal{U}}^\dagger D_a(\tilde{\mathbf{A}})\hat{\mathcal{U}},
		\end{equation}
		we get
		\begin{equation}\label{H-gauge}
			H_0(\mathbf{A},\tau)=\hat{\mathcal{U}}^\dagger H_0(\tilde{\mathbf{A}},\tau)\hat{\mathcal{U}}.
		\end{equation}
		Now we can see that the LLL wave function after the Dehn twist, when written in $(X^1,X^2)$, is 
		\begin{equation}\label{Landau-level-gauge}
			\hat{\mathcal{U}}\Psi_m(X'^1,X'^2|\tau+1) = \mathrm{e}^{i\pi\frac{m^2}{N_\phi}}\Psi_m(X^1,X^2|\tau), 
		\end{equation}
		where we have used 
		\begin{eqnarray}
			\nonumber
			&&\theta_{\frac{m}{N_\phi}}(N_{\phi}z|N_{\phi}(\tau+1)) \nonumber\\
			&=& \sum_n \mathrm{e}^{i\pi N_{\phi} \tau (n+\frac{m}{N_{\phi}})^2 + i2 \pi (n+\frac{m}{N_{\phi}})N_{\phi}z 
				+i\pi N_{\phi} (n+\frac{m}{N_{\phi}})^2} \nonumber\\
			&=& e^{i\pi \frac{m^2}{N_{\phi}}}\sum_n (-1)^{nN_\phi}\mathrm{e}^{i\pi N_{\phi} \tau (n+\frac{m}{N_{\phi}})^2 + i2 \pi (n+\frac{m}{N_{\phi}})N_{\phi}z}\nonumber \\
			&=& e^{i\pi \frac{m^2}{N_{\phi}}}\theta_{\frac{m}{N_\phi}}(N_{\phi}z|N_{\phi}\tau).
		\end{eqnarray} 
		Therefore, a phase factor $\mathrm{e}^{i\pi\frac{m^2}{N_\phi}}$ is gained in the $m$th LLL orbital after the Dehn twist.
		
		Similar to the $\mathcal{S}$-transformation, we introduce a many-body gauge transformation
		\begin{equation}\label{many-body-gauge}
			\hat{\mathcal{U}}_g = \exp{\left\lbrace i\pi N_{\phi} \sum_{I,i}\left( \frac{y_i^I}{L\tau_2}\right)^2\right\rbrace}
		\end{equation}
		to relate the many-body wave functions before and after the modular transformation $\left\lbrace \mathcal{T}:\tau \rightarrow \tau + 1 \right\rbrace $.
		Using Eqs.~(\ref{Abelian-Wave_append})-(\ref{Normalization}), we can get
		\begin{eqnarray}\label{tau+1-wave}
			\nonumber
			\hat{\mathcal{U}}_g\Psi^{\bm{a}}\left(\left\lbrace z_i^I \right\rbrace| \tau+1\right)
			&=&  \hat{\mathcal{U}}_g\mathcal{N}(\tau+1) f_c^{\bm a}(\left\lbrace Z^I \right\rbrace|\tau+1 ) f_r( \left\lbrace z_i^I \right\rbrace|\tau+1 ) 
			\mathrm{e}^{-i\pi N_{\phi} (\tau+1) \sum_{I,i}\left( y_i^I/L\tau_2\right)^2} \\
			\nonumber
			&=&  \mathcal{N}(\tau) f_c^{\bm a}(\left\lbrace Z^I \right\rbrace|\tau ) f_r( \left\lbrace z_i^I \right\rbrace|\tau ) 
			\mathrm{e}^{i\pi N_{\phi} \tau \sum_{I,i}\left( y_i^I/L\tau_2\right)^2}
			\mathrm{e}^{\frac{1}{12}i\pi (\bm{N}^T\cdot\bm{N}-\kappa )} \mathrm{e}^{i2\pi h_{\bm{\bm{\alpha}}}} \\
			&=& \Psi^{\bm{a}}\left(\left\lbrace X^{1}_{I,i},X^{2}_{I,i} \right\rbrace| \tau\right) \mathrm{e}^{\frac{1}{12}i\pi (\bm{N}^T\cdot\bm{N}-\kappa )} \mathrm{e}^{i2\pi h_{\bm{\bm{a}}}},
		\end{eqnarray}
		where $\bm N=\sum_I N^I \bm e_I$ is a vector enclosing the particle number in each component $N^I$ and we have used the following useful relations (here we assume the total flux $N_{\phi}$ through the torus is even):
		\begin{eqnarray}\label{Modular-T-Properties}
			\eta(\tau+1) &=& e^{i\pi/12}\eta(\tau),\nonumber\\
			\theta_{11}(z|\tau+1)    &=& e^{i\pi/4}\theta_{11}(z|\tau),\nonumber\\
			f^{(\bm{a})}(\bm{Z}|\tau+1)  &=&   e^{i\pi \bm{a}^T\cdot\bm{a}}f^{(\bm{a})}(\bm{Z}|\tau). \nonumber
		\end{eqnarray}
		Eq.~(\ref{tau+1-wave}) immediately gives the matrix representation of the $\mathcal{T}$-transform as
		\begin{equation}
			\mathcal{T}^{\bm{ab}} =\langle  \Psi^{\bm{b}};\tau|\mathcal{T}|\Psi^{\bm{a}};\tau\rangle = \delta_{\bm{a}\bm{b}}  \mathrm{e}^{i2\pi (h_{\bm{a}}-\frac{c}{24})} \mathrm{e}^{\frac{1}{12}i\pi \bm{N}^T\cdot\bm{N}},
		\end{equation}
		where $c=\kappa$ is the chiral central charge of the underlying edge CFT and $h_{\bm{a}}$ is the topological spin of the topological sector $\bm{a}$ satisfying
		\begin{eqnarray}\label{topological-spin}
			h_{\bm{a}}	&=&	\frac12 \bm{a}^T\cdot \bm{a}(\mathrm{mod}~1)
		\end{eqnarray}
		Again, we find that the result of $\mathcal T$ matrix is consistent with the calculation based on the CFT (Eq.~(6) in the main text), except for a particle number related phase factor.

		\section{III. Modular matrices for the Halperin $(m,m,m-1)$ state} 
		\label{SMC}
		
		In this section, we will explicitly calculate the modular matrices for the Halperin $(m,m,m-1)$ state, using the method of anyon condensation and extended chiral algebra, which serves as an example to validate our theory. We see that these direct calculations for the Halperin $(m,m,m-1)$ state give consistent results with those obtained in the previous sections. The difference is that, the calculation shown in this section is specific to $(m,m,m-1)$ state, while the CFT and trial wave function methods shown above are more general for any Halperin state. 
		
		\subsection{1. Anyon condensation method for the Halperin $(m,m,m-1)$ state}
		\label{append:singlet}
		Apart from the single-component Laughlin state, we consider another special class of FQH states called the Halperin $(m,m,m-1)$ state at filling $\nu=\frac{2}{2m-1}$. The Halperin state lives on a double-layered 
		system, where the particles are labeled by the pseudo-spin index $\{\uparrow,\downarrow\}$ and its wave function is (on the disk)~\cite{Halperin1983}
		\begin{align}\nonumber
			\Psi(\{z^\uparrow,z^\downarrow\})&= \prod_{i<j}(z_i^\uparrow-z_i^\uparrow)^m\prod_{i<j}(z_i^\downarrow-z_i^\downarrow)^m\prod_{i,j}(z_i^\uparrow-z_i^\downarrow)^{m-1}\\
			&\times \exp{\left[ -\frac14\sum_i \left( |z_i^\uparrow|^2+|z_i^\downarrow|^2 \right)  \right]}.
		\end{align}
		The bosonic Halperin $(m,m,m-1)$ state ($m$ is even) is described by the $\widehat{su(2)}_1$ WZW model with a $\widehat{u(1)}_{4m-2}$ boson~\cite{MOORE1991362}. 
		The central charge, topological spin and fusion rules are given by~\cite{yellowbook}
		\begin{align}
			&c=2,\qquad h_{(\lambda,a)} = \frac{\lambda(\lambda+2)}{12}+\frac{a^2}{8m-4},\\
			&(\lambda, a)\otimes(\mu, b) = \left( (\lambda+\mu)_{\mr{mod}~2},(a+b)_{\mr{mod}~4m-2}  \right) 
		\end{align}
		where $\lambda,\mu=0,1$ and $a,b=0,1,\cdots,4m-3$. Both fields have quantum dimension $\mr{d}_{(\lambda,a)}=\mr{d}_{\lambda}\mr{d}_{a}=1$. 
		
		Here, the number of primary fields $2\times(4m-2) = 8m-4$ is greater than the ground-state degeneracy $2m-1$ of the Halperin $(m,m,m-1)$ state on the torus. The reason can be understood by noticing a special primary 
		field with integer topological spin
		$ h_{(1,2m-1)}=\frac14+\frac{(2m-1)^2}{4(2m-1)}=\frac{m}{2}\in\mathbb{Z}$. The existence of such a bosonic field (with integer spin) beside vacuum indicates the phenomenon Bose-condensation~\cite{Slingerland_2009}. Let us denote this field as $J=(1,2m-1)$. The period of $J$ is $2$, since its fusion rule is $J^2=(0,0)$. $J$ should behave like the identity $\mathbf{1}=(0,0)$. The reason is, similar to the case that we cannot distinguish $(\lambda,a)$ from $(\lambda,a)\otimes \mathbf{1}^m$ ($m\in\mathbb{N}$), we cannot distinguish  $(\lambda,a)$ from $(\lambda,a)\otimes J^m$($m\in\mathbb{N}$) either. 
		After condensation, the $8m-4$ primary fields are mapped into $4m-2$ distinguishable primary fields $\{[(0,a)]_J|a=0,1,\cdots,4m-3 \}$, where $[\cdots]$ denotes a equivalence class under the fusion with $J$ field $[(0,a)]_J:(0,a)\sim(0,a)\otimes J$. For the notation simplicity, we will abbreviate $[(0,a)]_J$ to $a$ in the following. Moreover, not all fields in $[(0,a)]_J$ are available, since some of corresponding anyons are confined unless all fields in the same equivalence class have the same topological spin. 
		The terminology deconfined means the anyons can exist in the bulk. To find the deconfined fields, we need to consider the difference between the topological spin of $a$ and $a\otimes J$(in the sense of modulo 1):
		\begin{equation}\nonumber
			h_{a\otimes J}-h_a = \frac14+\frac{(a+2m-1)}{8m-4}-\frac{a^2}{8m-4}=\frac{a}{2}.
		\end{equation}
		Clearly, for even $a$ two anyons only differ by an integer spin. Thus only half anyons in set $[(0,a)]_J$ are deconfined and form the final theory
		\begin{equation}
			\{ a=[(0,2a)]_J|a=0,1,\cdots,2m-2 \}
		\end{equation}
		with the topological spins 
		\begin{equation}
			h_a = \frac{a^2}{2m-1},\qquad a\otimes b = (a+b)_{\mr{mod}2m-1},
		\end{equation}
		which gives the modular $\mathcal T$ matrix as
		\begin{align}
			\mathcal T_{ab} = \delta_{ab} e^{2\pi i \frac{a^2}{2m-1}}
		\end{align}
		Using the formula
		\begin{equation}
			\mathcal S_{ab} = \frac{1}{\mathcal{D}}\sum_c \mathcal{N}_{a \bar b}^{~~~c} \exp\left[2\pi i (h_c-h_a-h_b)\right] \mr{d}_c,
		\end{equation}
		the $\mathcal{S}$ matrix of the final theory is 
		\begin{equation}\label{res_CFT}
			\mathcal S_{ab} = \frac{1}{\sqrt{2m-1}}\exp\left(-2\pi i\frac{2ab}{2m-1} \right).
		\end{equation}
		The modular matrix obtained here exactly matches the results from the CFT and trial wave function. 

		\subsection{2. Extended chiral algebra method for the Halperin $(m,m,m-1)$ state}
		Furthermore, there is another way to construct the modular matrices, as we will illustrate below.
		The existence of an extra bosonic field $h_a=0\text{mod} 1$ signals the possibility of a ``block-diagonal modular invariant" partition function 
		with a ``extended vacuum block"~\cite{yellowbook}
		$$ Z=|\chi_0+\chi_{n_1}+\cdots+\chi_{n_i}|^2+\cdots .$$
		In our example $\mathcal{A}=\widehat{su(2)}_1\oplus\widehat{u(1)}_{2m-1}$, the characters of all primary fields are 
		\begin{equation} \label{eq:product_character}
			\chi_{(\lambda,a)}(\tau)=\chi^{\widehat{su(2)}_1}_\lambda(\tau)K_a^{(4m-2)}(\tau),
		\end{equation}
		where $
		\chi^{\widehat{su(2)}_1}_{\lambda}(\tau)=\frac{1}{\eta(\tau)}\sum_{m\in\mathbb{Z}+\lambda/2}q^{m^2}=
		K^{(2)}_{\lambda}(\tau)$ is the character of $\widehat{su(2)}_1$ and it is equivalent to the $\widehat{u(1)}_2$. The bosonic field $h_{(1,2m-1)}=1$ extends the vacuum sector to ``vacuum block" $C_0(\tau)=\chi_{(0,0)}(\tau)+\chi_{(1,2m-1)}(\tau)$, and other ``blocks" contain fields with topological spin differing by integers
		$$ C_p(\tau) = \chi_{(0,2p)}(\tau)+\chi_{(1,2p+2m-1)}(\tau).$$
		In terms of the block character $C_p(\tau)$, the block-diagonal modular invariant partition function can be written as
		$$ Z=\sum_{p=0}^{2m-2} |C_p(\tau)|^2. $$
		The integer difference of primary fields ensures the modular-$\mathcal{T}$ invariant, and the $\mathcal{S}$ transformation of block characters is (see Eqs.~(\ref{eq:extendedS}) and (\ref{eq:product_character}))
		\begin{align}\nonumber
			&C_p(-1/\tau)=\\\nonumber&\frac{1}{\sqrt{2m-1}}\sum_{q=0}^{2m-2} 
			\left[ e^{-2\pi i\frac{2pq}{2m-1}}\chi_{0,2q}(\tau)+e^{-2\pi i\frac{2p(2q+1)}{2m-1}}\chi_{1,2q+1}(\tau) \right]\\\nonumber
			&=\sum_{q=0}^{2m-2} \frac{e^{-2\pi i\frac{2pq}{2m-1}}}{\sqrt{2m-1}}
			\left[ \chi_{0,2q}(\tau)+\chi_{1,2q+2m-1}(\tau) \right]\\\nonumber
			&=\sum_{q=0}^{2m-2} \frac{e^{-2\pi i\frac{2pq}{2m-1}}}{\sqrt{2m-1}}C_q(\tau).
		\end{align}
		Therefore, the $\mathcal{S}$ matrix of the block characters is
		\begin{equation}
			\mathcal{S}_{pq} = \frac{e^{-2\pi i\frac{2pq}{2m-1}}}{\sqrt{2m-1}},
		\end{equation} 
		which is consistent with Eq.~(\ref{res_CFT}).
		The unitary $\mathcal{S}^\dagger\mathcal{S}=1$ gives the modular-$\mathcal{S}$ invariant of the partition function $Z$. Conclusively, the bosonic field gives another type of modular invariant, the block-diagonal modular invariant, which can be constructed by grouping the fields with integer difference topological spin into blocks. After that, the modular $\mathcal{S}$ matrix of the block characters is the final result that we are looking for. Meanwhile, the structure of the vacuum block indicates a symmetry enhancement or an extended chiral algebra, which is the illustration of anyon condensation.

		\section{IV. More examples}
		\label{SMD}
		Our theory is general, which has broad applications. 
		In this section, we present some examples of modular matrices of multi-component states that are beyond two-component spin-singlet states.

		\subsection{1. Halperin $(441)$ state}
		\label{SMD1}
\begin{figure}[t]
			\centering
			\includegraphics[width=0.45\linewidth]{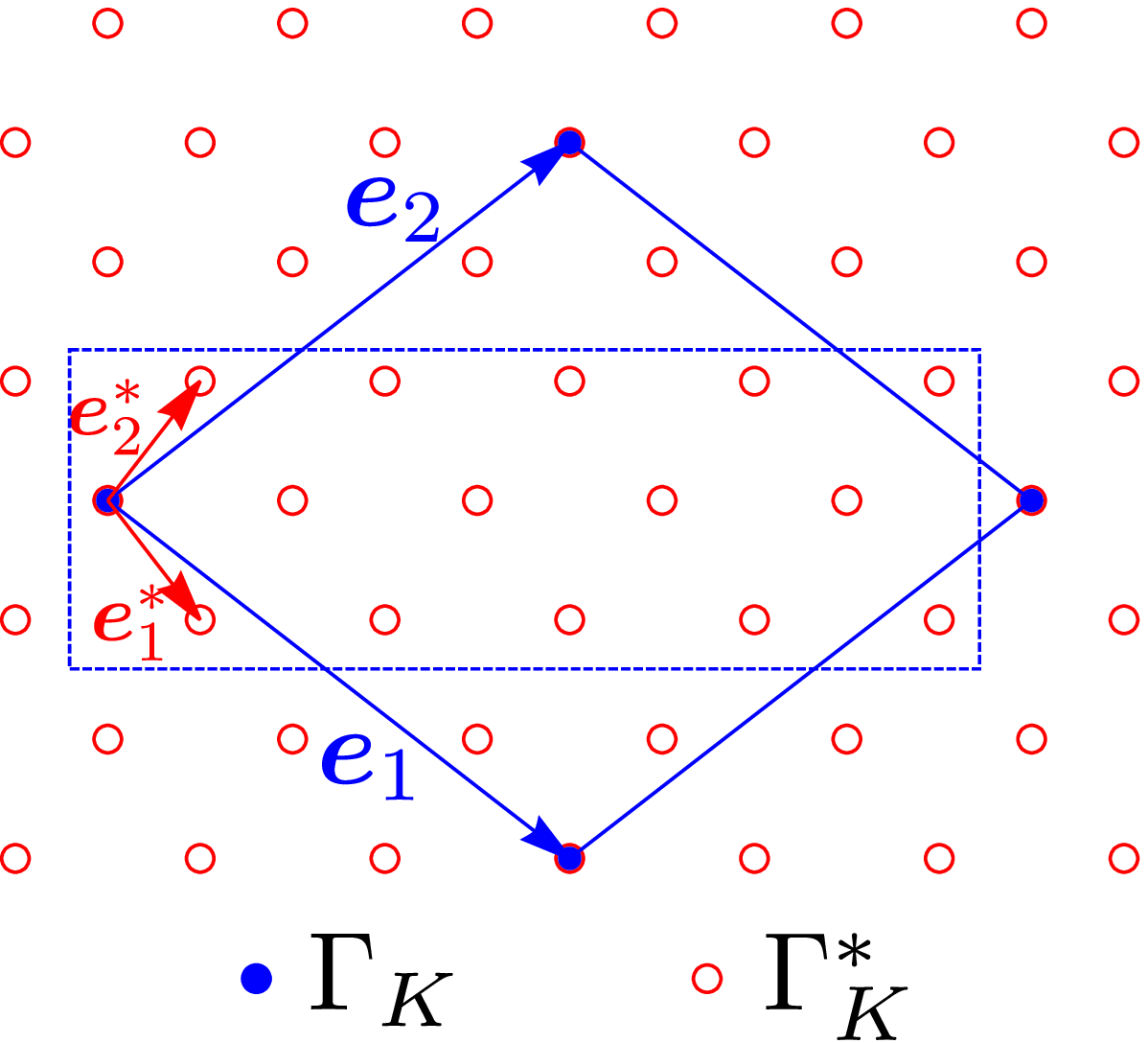} 
			\caption{\label{fig:441}
				The $\Gamma_K$ and $\Gamma_K^*$ lattices for the $(441)$ state. The solid blue parallelogram is the coset of $(441)$ states. An equivalent representation of coset is shown by the dashed blue rectangular.
			}
		\end{figure}

		This subsection gives an example of a double-layer $(441)$ state whose $K$ matrix is 
		\begin{equation}
			K=\left( \begin{array}
				{cc} 4&1\\1&4
			\end{array} \right).
		\end{equation}
		This filling of this state is $\nu=2/5$, and the degeneracy $D=|\det K|=15=5\times 3$. Numerically, this state has been found recently~\cite{TSZeng2022}. The K-lattice and its dual lattice are shown in Fig.~\ref{fig:441}. The coset is written as $\Gamma_K^*/\Gamma_K = \left\{\Gamma_0+a\bm e^*_1+a\bm e^*_2 | a=0,1,2,3,4 \right\}$ where $\Gamma_0=\{ (0,0),\bm e^*_1, \bm e^*_2 \}$. According to the results of the modular matrix [Eqs.~(4) and (6) in the main text)], we obtain 
		
		\begin{equation}  
			\mathcal{T} = e^{-2\pi(\frac{2}{24})}\mathrm{diag} 
			\left\{0,\frac{2}{15},\frac{2}{15},\frac{1}{5},\frac{8}{15},\frac{8}{15},\frac{4}{5},\frac{1}{3},\frac{1}{3},\frac{4}{5},\frac{8}{15},\frac{8}{15},\frac{1}{5},\frac{2}{15},\frac{2}{15}\right\}\\
		\end{equation}
		and
		\begin{equation}
			\mathcal{S} = \frac{1}{\sqrt{15}}
			\left(
			\begin{array}{ccccc}
				\left(
				\begin{array}{ccc}
					1 & 1 & 1 \\
					1 & \kappa ^4 & \frac{1}{\kappa } \\
					1 & \frac{1}{\kappa } & \kappa ^4 \\
				\end{array}
				\right) & \left(
				\begin{array}{ccc}
					1 & 1 & 1 \\
					\kappa ^3 & \kappa ^7 & \kappa ^2 \\
					\kappa ^3 & \kappa ^2 & \kappa ^7 \\
				\end{array}
				\right) & \left(
				\begin{array}{ccc}
					1 & 1 & 1 \\
					\kappa ^6 & \kappa ^{10} & \kappa ^5 \\
					\kappa ^6 & \kappa ^5 & \kappa ^{10} \\
				\end{array}
				\right) & \left(
				\begin{array}{ccc}
					1 & 1 & 1 \\
					\kappa ^9 & \kappa ^{13} & \kappa ^8 \\
					\kappa ^9 & \kappa ^8 & \kappa ^{13} \\
				\end{array}
				\right) & \left(
				\begin{array}{ccc}
					1 & 1 & 1 \\
					\kappa ^{12} & \kappa ^{16} & \kappa ^{11} \\
					\kappa ^{12} & \kappa ^{11} & \kappa ^{16} \\
				\end{array}
				\right) \\
				\left(
				\begin{array}{ccc}
					1 & \kappa ^3 & \kappa ^3 \\
					1 & \kappa ^7 & \kappa ^2 \\
					1 & \kappa ^2 & \kappa ^7 \\
				\end{array}
				\right) & \left(
				\begin{array}{ccc}
					\kappa ^6 & \kappa ^9 & \kappa ^9 \\
					\kappa ^9 & \kappa ^{16} & \kappa ^{11} \\
					\kappa ^9 & \kappa ^{11} & \kappa ^{16} \\
				\end{array}
				\right) & \left(
				\begin{array}{ccc}
					\kappa ^{12} & \kappa ^{15} & \kappa ^{15} \\
					\kappa ^{18} & \kappa ^{25} & \kappa ^{20} \\
					\kappa ^{18} & \kappa ^{20} & \kappa ^{25} \\
				\end{array}
				\right) & \left(
				\begin{array}{ccc}
					\kappa ^{18} & \kappa ^{21} & \kappa ^{21} \\
					\kappa ^{27} & \kappa ^{34} & \kappa ^{29} \\
					\kappa ^{27} & \kappa ^{29} & \kappa ^{34} \\
				\end{array}
				\right) & \left(
				\begin{array}{ccc}
					\kappa ^{24} & \kappa ^{27} & \kappa ^{27} \\
					\kappa ^{36} & \kappa ^{43} & \kappa ^{38} \\
					\kappa ^{36} & \kappa ^{38} & \kappa ^{43} \\
				\end{array}
				\right) \\
				\left(
				\begin{array}{ccc}
					1 & \kappa ^6 & \kappa ^6 \\
					1 & \kappa ^{10} & \kappa ^5 \\
					1 & \kappa ^5 & \kappa ^{10} \\
				\end{array}
				\right) & \left(
				\begin{array}{ccc}
					\kappa ^{12} & \kappa ^{18} & \kappa ^{18} \\
					\kappa ^{15} & \kappa ^{25} & \kappa ^{20} \\
					\kappa ^{15} & \kappa ^{20} & \kappa ^{25} \\
				\end{array}
				\right) & \left(
				\begin{array}{ccc}
					\kappa ^{24} & \kappa ^{30} & \kappa ^{30} \\
					\kappa ^{30} & \kappa ^{40} & \kappa ^{35} \\
					\kappa ^{30} & \kappa ^{35} & \kappa ^{40} \\
				\end{array}
				\right) & \left(
				\begin{array}{ccc}
					\kappa ^{36} & \kappa ^{42} & \kappa ^{42} \\
					\kappa ^{45} & \kappa ^{55} & \kappa ^{50} \\
					\kappa ^{45} & \kappa ^{50} & \kappa ^{55} \\
				\end{array}
				\right) & \left(
				\begin{array}{ccc}
					\kappa ^{48} & \kappa ^{54} & \kappa ^{54} \\
					\kappa ^{60} & \kappa ^{70} & \kappa ^{65} \\
					\kappa ^{60} & \kappa ^{65} & \kappa ^{70} \\
				\end{array}
				\right) \\
				\left(
				\begin{array}{ccc}
					1 & \kappa ^9 & \kappa ^9 \\
					1 & \kappa ^{13} & \kappa ^8 \\
					1 & \kappa ^8 & \kappa ^{13} \\
				\end{array}
				\right) & \left(
				\begin{array}{ccc}
					\kappa ^{18} & \kappa ^{27} & \kappa ^{27} \\
					\kappa ^{21} & \kappa ^{34} & \kappa ^{29} \\
					\kappa ^{21} & \kappa ^{29} & \kappa ^{34} \\
				\end{array}
				\right) & \left(
				\begin{array}{ccc}
					\kappa ^{36} & \kappa ^{45} & \kappa ^{45} \\
					\kappa ^{42} & \kappa ^{55} & \kappa ^{50} \\
					\kappa ^{42} & \kappa ^{50} & \kappa ^{55} \\
				\end{array}
				\right) & \left(
				\begin{array}{ccc}
					\kappa ^{54} & \kappa ^{63} & \kappa ^{63} \\
					\kappa ^{63} & \kappa ^{76} & \kappa ^{71} \\
					\kappa ^{63} & \kappa ^{71} & \kappa ^{76} \\
				\end{array}
				\right) & \left(
				\begin{array}{ccc}
					\kappa ^{72} & \kappa ^{81} & \kappa ^{81} \\
					\kappa ^{84} & \kappa ^{97} & \kappa ^{92} \\
					\kappa ^{84} & \kappa ^{92} & \kappa ^{97} \\
				\end{array}
				\right) \\
				\left(
				\begin{array}{ccc}
					1 & \kappa ^{12} & \kappa ^{12} \\
					1 & \kappa ^{16} & \kappa ^{11} \\
					1 & \kappa ^{11} & \kappa ^{16} \\
				\end{array}
				\right) & \left(
				\begin{array}{ccc}
					\kappa ^{24} & \kappa ^{36} & \kappa ^{36} \\
					\kappa ^{27} & \kappa ^{43} & \kappa ^{38} \\
					\kappa ^{27} & \kappa ^{38} & \kappa ^{43} \\
				\end{array}
				\right) & \left(
				\begin{array}{ccc}
					\kappa ^{48} & \kappa ^{60} & \kappa ^{60} \\
					\kappa ^{54} & \kappa ^{70} & \kappa ^{65} \\
					\kappa ^{54} & \kappa ^{65} & \kappa ^{70} \\
				\end{array}
				\right) & \left(
				\begin{array}{ccc}
					\kappa ^{72} & \kappa ^{84} & \kappa ^{84} \\
					\kappa ^{81} & \kappa ^{97} & \kappa ^{92} \\
					\kappa ^{81} & \kappa ^{92} & \kappa ^{97} \\
				\end{array}
				\right) & \left(
				\begin{array}{ccc}
					\kappa ^{96} & \kappa ^{108} & \kappa ^{108} \\
					\kappa ^{108} & \kappa ^{124} & \kappa ^{119} \\
					\kappa ^{108} & \kappa ^{119} & \kappa ^{124} \\
				\end{array}
				\right) \\
			\end{array}
			\right)
		\end{equation}
		with $\kappa=e^{-2\pi i\frac{1}{15}}$. Each small $3\times3$ matrix corresponds to the three points in $\Gamma_0+a(\bm e_1+\bm e_2)=\{a\bm e_1+a\bm e_2,(a+1)\bm e_1+a\bm e_2,a\bm e_1+(a+1)\bm e_2\}$.

		We would like to stress that, here the representation of the coset $\Gamma_K^*/\Gamma_K $ of the Halperin $(441)$ state
		is two-dimensional, which is different from the one-dimensional representation of the spin-singlet state (see Fig.~1 in the main text).

		\subsection{2. Halperin $(m,m,m,m-1)$ state}
		\label{SMD2}
\begin{figure}[t]
			\centering
			\includegraphics[width=0.45\linewidth]{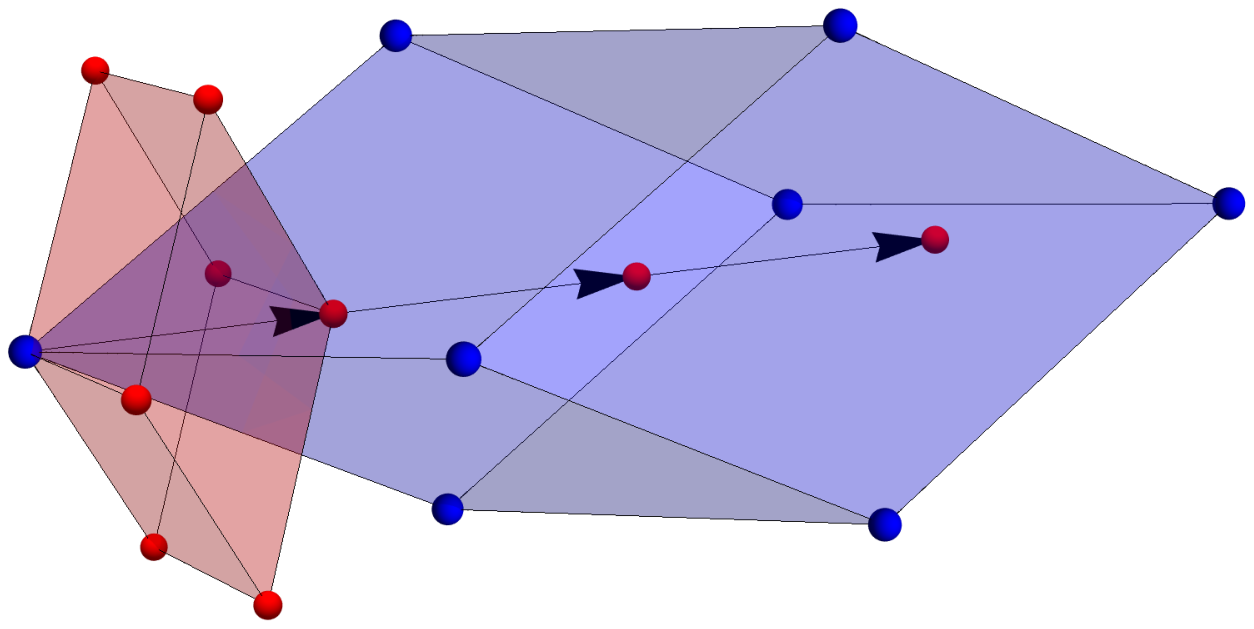} 
			\includegraphics[width=0.45\linewidth]{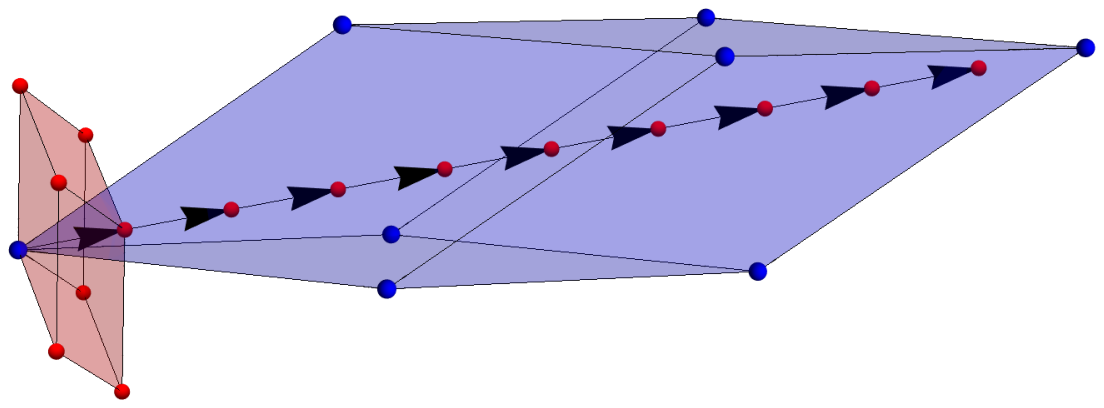}	
			\caption{\label{fig:2221}
				The $\Gamma_K$ and $\Gamma_K^*$ lattices for (left) $(2221)$ and (right) $(4443)$ states. The blue and red spheres represent the lattice points of $\Gamma_K$ and $\Gamma_K^*$, respectively. The red spheres inside the parallelepiped are the coset lattice of $\Gamma_K^*/\Gamma_K$.
			}
		\end{figure}

		In this subsection we consider the three-component Halperin $(m,m,m,m-1)$ state. The $K$ matrix of $(m,m,m,m-1)$ state is defined as
		\begin{equation}
			K=\left( \begin{array}
				{ccc}
				m&m-1&m-1\\m-1&m&m-1\\m-1&m-1&m
			\end{array} \right)
		\end{equation}
		with filling $\nu=3/(3m-2)$ and degeneracy $D=|\det K|=3m-2$. Two examples of the $K$-lattice and its dual lattice are shown in Fig.~\ref{fig:2221}. The coset lattice is $\Gamma_K^*/\Gamma_K=\{ \frac{1}{3m-2}(\bm e_1+\bm e_2+\bm e_3) |a=0,1,\cdots 3m-3\}$, and we can write down their modular matrices accordingly:
		\begin{equation}
			\mathcal{T}_{aa} = e^{2\pi i(\frac{3a^2}{6m-4}-\frac{3}{24})},\qquad
			\mathcal{S}_{ab} = \frac{1}{\sqrt{3m-2}} e^{ -6\pi i \frac{ab}{3m-2}}.
		\end{equation}
		As an example, if we choose $m=2$,  the modular matrices are
		\begin{eqnarray}
			\mathcal{T} &= e^{-\frac{\pi i}{4}}\left(
			\begin{array}{cccc}
				1 & 0 & 0 & 0 \\
				0 & e^{2\pi i\frac{3}{8}} & 0 & 0 \\
				0 & 0 & e^{2\pi i\frac{1}{2}} & 0 \\
				0 & 0 & 0 & e^{2\pi i\frac{3}{8}} \\
			\end{array}
			\right), \\
			\mathcal{S} &= \frac{1}{2}\left(
			\begin{array}{cccc}
				1 & 1 & 1 & 1 \\
				1 & i & -1 & -i \\
				1 & -1 & 1 & -1 \\
				1 & -i & -1 & i \\
			\end{array}
			\right).
		\end{eqnarray}

	\end{document}